\documentclass[journal,10pt]{IEEEtran}

\usepackage{algorithm,multirow}
\usepackage{algorithmic}
\usepackage{graphicx}
\usepackage{comment}
\usepackage{cite}
\usepackage{multirow}
\usepackage{amssymb,amsthm,mathrsfs}
\usepackage{amsfonts,epsf,graphicx,times,amsmath,multirow,url,cite}
\usepackage{float}
\usepackage{flexisym}
\usepackage[capitalise]{cleveref}

\usepackage{mathtools}

\usepackage{amsthm}
\usepackage{subcaption}
\usepackage{color}
\usepackage{amsmath}

\graphicspath{{figures/}}
 \usepackage{url}
\newcommand{\sol}{IQuCS}

\newcommand{\soll}{RESAQuS}

\begin{document}

\title{Resource-Efficient and Self-Adaptive Quantum Search in a Quantum-Classical Hybrid System}

\author{
Zihao Jiang*, Zefan Du*\thanks{*Authors contributed equally}, Shaolun Ruan\\  
Juntao Chen, Yong Wang,  Long Cheng, 	Rajkumar Buyya, and Ying Mao
\IEEEcompsocitemizethanks{

\IEEEcompsocthanksitem Z. Jiang, Z. Du, J. Chen and Y. Mao are with the Department of Computer and Information Science, Fordham University, New York City. E-mail: \{zjiang86, zdu19, jchen504, ymao41\}@fordham.edu
\IEEEcompsocthanksitem S. Ruan and Y. Wang are with the School of Computing and Information Systems,  Singapore Management University. E-mail: \{slruan.2021, yongwang\}@smu.edu.sg
\IEEEcompsocthanksitem L. Cheng is with the School of Control and Computer Engineering, North China Electric Power University, Beijing, China. E-mail: lcheng@ncepu.edu.cn
\IEEEcompsocthanksitem R. Buyya is with the School of Computing and Information Systems, The University of Melbourne, Australia. E-mail: rbuyya@unimelb.edu.au
}
}


\maketitle
\begin{abstract}

Over the past decade, the rapid advancement of deep learning and big data applications has been driven by vast datasets and high-performance computing systems. However, as we approach the physical limits of semiconductor fabrication in the post-Moore's Law era, questions arise about the future of these applications. In parallel, quantum computing has made significant progress with the potential to break limits. Major companies like IBM, Google, and Microsoft provide access to noisy intermediate-scale quantum (NISQ) computers. Despite the theoretical promise of Shor's and Grover's algorithms, practical implementation on current quantum devices faces challenges, such as demanding additional resources and a high number of controlled operations. To tackle these challenges and optimize the utilization of limited onboard qubits, we introduce ReSaQuS, a resource-efficient index-value searching system within a quantum-classical hybrid framework. Building on Grover's algorithm, ReSaQuS employs an automatically managed iterative search approach. This method analyzes problem size, filters less probable data points, and progressively reduces the dataset with decreasing qubit requirements. Implemented using Qiskit and evaluated through extensive experiments, ReSaQuS has demonstrated a substantial reduction, up to 86.36\% in cumulative qubit consumption and 72.72\% in active periods, reinforcing its potential in optimizing quantum computing application deployment.

\end{abstract}


\maketitle

\section{Introduction}


Over the past decade, we have witnessed groundbreaking advancements in deep-learning and big-data based applications. New algorithms, coupled with increased computational power and modern design methodologies, have facilitated a broad spectrum of applications, ranging from scientific data processing to commercial image and speech recognition. At the backend side, these developments are underpinned by massive datasets and powered by high-performance computing systems with substantial computational resources. However, in the post-Moore's Law era, we are confronted by the looming physical limitations of semiconductor fabrication. Combined with the ceaseless growth of data volumes, this circumstance compels us to reassess the future of these applications. 

To navigate the challenges of this new era, researchers are turning to emerging disciplines. Quantum computing 
provides a promising alternative. 
For example, Google exhibited quantum supremacy with a 53-qubit quantum computer, completing in 200 seconds a task projected to take 10,000 years on the world's largest classical computer. Given the potential, quantum-based applications are gaining increasing attentions in both industry and academia.

Inherently rooted in quantum mechanics, quantum bits (qubits) offer distinctive characteristics such as superposition and entanglement. Unlike their classical counterparts' deterministic states, a qubit in superposition simultaneously embodies both the '0' and '1' states, significantly enhancing its representative capacity. Furthermore, two qubits can be placed in an entanglement state, meaning the measurement of one qubit is contingent on the measurement of another.

Leveraging these distinctive properties, many algorithms have been proposed, aiming to achieve quantum speedups. Among them, Grover's algorithm stands as a representative example. It targets on unstructured database search and outperforms classical counterparts, yielding a quadratic speedup. 
With Grover's algorithm, the qubits start out in the uniform superpositions such that the amplitudes of all data points are the same. Then, it utilizes an oracle function $O$, a "black box" function that only reflects the amplitudes of the searching targets and remains others untouched. 
Next, the algorithm applies another reflection that can amplify the amplitude of the searching targets and deamplify others. With certain rounds of this amplitude amplification process, the targets will have significantly higher amplitudes compared to others. 
Each reflection is an invocation of Grover's operator. While classical unstructured searching algorithms operate with an $O(N)$ complexity, Grover's algorithm impressively trims it down to $O(\sqrt{N})$. 

Grover's search algorithm, along with its generalized version Quantum Amplitude Estimation (QAE), has garnered significant interest due to its genericity and quadratic speedup, serving as a cornerstone for myriad quantum-based applications~\cite{grassl2016applying, zhang2020depth, byrnes2018generalized, morales2018variational, orus2019quantum, plekhanov2022variational, ramezani2020machine}. Brassard {\em et al.}~\cite{brassard1998quantum} for instance, have proposed a quantum counting algorithm that combines Grover's and Shor's~\cite{shor1999polynomial} algorithms to count the number of targets in a given dataset. This can be perceived as an implementation of QAE based on Quantum Phase Estimation (QPE).
However, despite Grover's search and QAE's potential for substantial speedups, its main component, QPE, grapples with two major obstacles: the requirement for extra qubits and numerous controlled operations. These challenges render QPE impractical in the Noisy Intermediate-Scale Quantum (NISQ) era, characterized by limited quantum resources, including qubits~\cite{suzuki2020amplitude, Wie19, aaronson2020quantum, grinko2021iterative}.


Optimized algorithms have been suggested to bypass QPE's reliance~\cite{suzuki2020amplitude, Wie19, aaronson2020quantum, grinko2021iterative}, potentially enhancing the efficiency of quantum counting and searching on NISQ quantum computers~\cite{rao2020quantum, willsch2020benchmarking, amico2019experimental, pelofske2021sampling}. However, these solutions maintain the same problem size in each Grover’s search iteration and fails to reduce the resources, e.g., qubits, requirement. 
In the literature, IQuCS~\cite{mu2022iterative} successfully reduces the search size iteratively by using classical data post-posting. However, its algorithms rely on an administrator-specified threshold that lacks adaptability in different search contexts.

The address the challenges, we introduce \soll~ that considers an unstructured dataset consisting of \textit{(index, value)} pairs. The input data from clients are values, and the index is the unstructured number that represents data.  \soll~ utilizes Grover's algorithm to search for targeted data points. 
However, instead of finishing the entire search with multiple repeated invocations, \soll~ sets breakpoints to divide the entire search into multiple iterations. Each iteration is an individual Grover's search with a reduced number of invocations. Based on the analysis of results after each iteration, \soll~filters out non-target pairs and uses the remaining data for the next iteration. The reduced dataset results in fewer qubits required. Furthermore, on the classical part of the system, \soll~ adopts a self-adaptive clustering method for data filtering. Additionally, \soll~updates the list of indexes and reconstructs the values iteratively, which further reduces the number of required qubits. 
The following summarizes the main contributions of this paper:

\begin{itemize}
    \item We propose, \soll, a resource-efficient and self-adaptive quantum search algorithm in a quantum-classical hybrid system for unstructured dataset of {\em (index, value)} pairs. 
    
    \item In \soll, our self-adaptive algorithms re-generates indexes and values to achieve resource efficiency. Meanwhile, it maintains the mapping from the generated data points to original ones and returns the original searching {\em (index, value)} pairs.   
    
    \item Furthermore, \soll~ utilizes a self-adaptive clustering algorithm analysis the  analyze quantum state fidelities after each iteration and filter out the non-target data points without finishing the entire search. The self-adaptive nature accommodate the searches with different problem sizes. 
    
    \item We implement \soll~ with Qiskit, a popular quantum programming framework. Intensive experiments are conducted with both Qiskit cloud simulator and experiments on IBM-Q. Additionally, we propose Cumulative-Qubit Consumption (CQC) to evaluate the qubit resource usage. The results demonstrate that \soll~ reduced CQC by up to 86.36\% comparing to the state-of-the-arts. Additionally, it reduces active periods by up to 72.72\% on quantum  workers in the cluster mode. 
    
\end{itemize}

The rest of this paper is organized as follows. Section II discusses the related works. Section III introduces the basics of quantum computing and Grover's algorithm. We present our \soll~ system design and algorithms in Section IV. Comprehensive evaluations of the developed \soll~ are provided in Section V. Section VI concludes the work.

\section{Related Work}

As the physical limits of semiconductor fabrication draw near, the race for quantum computing development is intensifying. Major players and emerging startups, including IBM, Google, Amazon, and IonQ, have initiated public access to quantum systems.


In the quantum landscape, foundational algorithms, such as Grover's algorithm~\cite{grover1996fast}, Shor's algorithm~\cite{shor1994algorithms}, quantum phase estimation~\cite{d1998general}, and variational quantum circuits~\cite{cerezo2021variational}, attract tremendous attention that aims to observe a quantum speedup on real devices in various applications. While these algorithms, in theory, can deliver quadratic or even exponential speedups~\cite{babbush2021focus, baheri2022pinpointing, ambainis2020quadratic, hoefler2023disentangling}, their practical deployment on NISQ devices presents formidable challenges. Quantum phase estimation, for instance, requires additional qubits and a substantial number of controlled operations, which are currently unfeasible due to the constraints of low-qubit count and noise in quantum hardware.

Building on these fundamental quantum components, advancements have been observed in the realms of quantum deep learning ~\cite{stein2022quclassi, stein2022qucnn, stein2021qugan, schutt2019unifying, l2024quantum, hoefler2023disentangling, alchieri2021introduction, yang2022semiconductor, ajagekar2021quantum, d2023distributed}, quantum visualization~\cite{ gunasekaran2020visualizing, ruan2023quantumeyes, li2022quantum, ruan2023venus, liu2022visualizing, ruan2022vacsen} and big data analytics~\cite{das2022experimental,baheri2021tqea, ablayev2019quantum, ding2021quantum}. Despite these improvements, the high quantum resource requirements remains a significant hurdle to achieving commercial deployment at scale in real-world scenarios. For instance, while QuGAN~\cite{stein2021qugan} and QuClassi~\cite{stein2022quclassi} demonstrate impressive performance in model building, their results are based on a merely 4-dimensional dataset on the IBM-Q platform. This is primarily due to the limitations of current NISQ hardware which offer a limited qubits and suffer from noise-related issues.

Various optimizations have been proposed to address the resource-intensive nature of QPE operations in quantum counting and searching, which hinders their practicality on NISQ machines~\cite{zhang2020depth, suzuki2020amplitude, Wie19, aaronson2020quantum, grinko2021iterative}. 
An exemplary approach is the depth optimization method presented in~\cite{zhang2020depth}. This approach harnesses multi-stage processing, along with global and local applications of Grover's operators, to achieve a substantial 20\% reduction in circuit depth.
Additionally, MLQAE~\cite{suzuki2020amplitude} combines multiple iterations of Grover's algorithm with maximum likelihood estimation to reduce the qubit requirement. Wie {\em et al.}~\cite{Wie19} proposes the use of Hadamard tests as less expensive alternatives to QPE. Another approach, introduced in~\cite{aaronson2020quantum}, presents a simplified quantum computing algorithm that operates without QPE but introduces a significant overhead. A recent effort called IQAE~\cite{grinko2021iterative} reduces the overhead by iteratively postprocessing quantum results using only Grover's operator. These optimized solutions have the potential to improve the efficiency of quantum counting and searching on NISQ computers~\cite{willsch2020benchmarking, pelofske2021sampling}. However, these optimizations often focus on specific problems or can be challenging to implement. Additionally, existing approaches treat algorithms as indivisible tasks with the same problem size in each iteration. In contrast, IQuCS~\cite{mu2022iterative} proposes an iterative approach that divides searching problems and effectively reduces the problem size using classical data post-processing. However, IQuCS relies on an administrator-specified threshold, which lacks adaptability in different search contexts and imposes additional deployment prerequisites.

Building upon IQuCS, our approach, \soll, tackles a quantum search problem that deals with {\em (index, value)} pairs. Diverging from existing literature, \soll~ places emphasis on iteratively reducing the input dataset by filtering out non-target elements. This reduction is achieved through quantum state fidelity analysis performed on the classical part. Furthermore, we introduce self-adaptive algorithms to accommodate diverse searching settings. By progressively reducing the input data in each iteration, \soll~ maximizes the utilization of qubits, resulting in an efficient completion of the search task.

\section{Background and Motivation}

The foundational unit in classical computing is the bit, which represents a logical state with possible values $1$ and $0$. Classical computing commonly implements bits using electromagnetic phenomena. 

The theory of quantum mechanics describes the physical properties of nature at atomic and subatomic scales, providing the foundations for understanding quantum computing. Two fundamental phenomena in the theory are superposition and entanglement, which form the building blocks in this field. Quantum bits, or qubits, are capable of realizing these phenomena.
Unlike classical bits, a qubit can simultaneously represent $0$ and $1$ in the superposition. It collapses to a deterministic state, $0$ or $1$, upon measurement.

\subsection{Qauntum States}

Orthogonal vectors can represent the states of a qubit. 
Equation~\ref{1} represents a qubit's 0 and 1 state.
\begin{equation} 
    |0\rangle=
    \begin{bmatrix}
    1\\
    0
    \end{bmatrix},
    \qquad
    |1\rangle=
    \begin{bmatrix}
    0\\
    1
    \end{bmatrix} 
\label{1}    
\end{equation}

A universal qubit's state $|q\rangle$ can be written as Equation~\ref{2},
\begin{equation}
    |q\rangle=\alpha|0\rangle+\beta|1\rangle,
\label{2}    
\end{equation}
where
\begin{equation}
    |\alpha|^2+|\beta|^2=1
\label{3}
\end{equation}
The vector $|q\rangle$ is called a state vector of qubits, representing the linear combination or superposition of two states associated with qubits ~\cite{thomas2022introduction}.
In Equation~\ref{3}, $\alpha$ and $\beta$, the amplitudes associated with each state, can also be interpreted as the probability for the qubit to be in states $|0\rangle$ and $|1\rangle$, respectively. When the qubit’s state is measured, these amplitudes represent the probability of the superposition collapsing into each state. Equation 3 is derived from the normalization requirement, which ensures the probability of a qubit being in any state.




\subsection{Quantum Gates}
Quantum computers encode and manipulate data using quantum gates. The two main types of quantum gates are rotation gates, which perform a rotation about an axis, and controlled gates, which operate on a qubit depending on the value of a control qubit. 

{\noindent\bf Rotation Gates}:
Rotation gates enable qubit rotations by adjusting parameters. The matrix representation of the generalized single-rotation gate $R$ is presented in Equation~\ref{eq:rotation_general}.
\begin{equation}
R(\theta,\phi)=\left[\begin{array}{cc}
\cos \frac{\theta}{2} & -ie^{-i\phi}\sin \frac{\theta}{2} \\
-ie^{-i\phi}\sin \frac{\theta}{2} & \cos \frac{\theta}{2}
\end{array}\right].
\label{eq:rotation_general}
\end{equation}


Three commonly used special cases of the general rotation gate are the $R_X$, $R_Y$, and $R_Z$ gates, as shown in \eqref{eq:rx}-\eqref{eq:rz}. These gates represent rotations in the $x$, $y$, and $z$ planes, respectively. They can be expressed as follows:
$R_X(\theta)$: This gate represents a rotation about the $x$-axis by an angle $\theta$. It is a special case of the general rotation gate, where $\phi = 0$.
$R_Y(\theta)$: This gate represents a rotation about the $y$-axis by an angle $\theta$. It is another special case of the general rotation gate, where $\phi = \frac{\pi}{2}$.
$R_Z(\theta)$: This gate represents a rotation about the $z$-axis by an angle $\theta$. The derivation of $R_Z$ from the general rotation gate is more involved and is not included here.
These special cases provide specific rotations around the corresponding axes, enabling precise control over the quantum state during quantum computations.

\begin{equation}
R_{X}(\theta)=\left[\begin{array}{cc}
\cos \frac{\theta}{2} & -i\sin \frac{\theta}{2} \\
-i\sin \frac{\theta}{2} & \cos \frac{\theta}{2}
\end{array}\right]=R(\theta,0)
\label{eq:rx}
\end{equation}

\begin{equation}
R_{Y}(\theta)=\left[\begin{array}{cc}
\cos \frac{\theta}{2} & -\sin \frac{\theta}{2} \\
\sin \frac{\theta}{2} & \cos \frac{\theta}{2}
\end{array}\right]=R(\theta,\frac{\pi}{2})
\label{eq:ry}
\end{equation}

\begin{equation}
R_{Z}(\theta)=\left[\begin{array}{cc}
e^{\frac{-i\theta}{2}} & 0 \\
0&e^{\frac{-i\theta}{2}}
\end{array}\right]
\label{eq:rz}
\end{equation}

{\noindent Hadamard Gate}:
The Hadamard gate is a fundamental gate in quantum computation. It is a single-qubit gate that puts a qubit into the superposition. The gate can be represented by Equation~\ref{eq:hgate}: 
\begin{equation}
H=\frac{1}{\sqrt{2}}\left[\begin{array}{cc}
1 & 1 \\
1 & -1
\end{array}\right].
\label{eq:hgate}
\end{equation}
The coefficient 
$\frac{1}{\sqrt{2}}$ 
in \eqref{eq:hgate} accounts for the normalization of state amplitudes, ensuring that the sum of their squares is equal to 1. Each state has a probability of $\frac{1}{2}$ and an amplitude of $\frac{1}{\sqrt{2}}$.

{\noindent Two-Qubit Rotation Gates}:
There are operations that act as two-qubit rotations, applying an equal rotation to both qubits. These gates are described by the following equations \eqref{eq:rxx}--\eqref{eq:rzz}. Note that these gates are represented by $4\times4$ matrices, unlike the single-qubit gates that are $2\times2$ matrices. This is because, in a two-qubit gate, each qubit has two possible measurements, resulting in four possible outcomes ($|00\rangle$, $|01\rangle$, $|10\rangle$, $|11\rangle$), as opposed to the two outcomes seen in single-qubit gates.

\begin{small}
\begin{equation}
R_{XX}(\theta)=\left[\begin{array}{cccc}
\cos \frac{\theta}{2} & 0 & 0 & -i\sin \frac{\theta}{2}\\
0 & \cos \frac{\theta}{2} & -i\sin \frac{\theta}{2} & 0 \\
0 & -i\sin \frac{\theta}{2} & \cos \frac{\theta}{2} & 0 \\
-i\sin \frac{\theta}{2} & 0 & 0 & \cos \frac{\theta}{2}
\end{array}\right]
\label{eq:rxx}
\end{equation}
\end{small}

\begin{small}
\begin{equation}
R_{YY}(\theta)=\left[\begin{array}{cccc}
\cos \frac{\theta}{2} & 0 & 0 & i\sin \frac{\theta}{2}\\
0 & \cos \frac{\theta}{2} & -i\sin \frac{\theta}{2} & 0 \\
0 & -i\sin \frac{\theta}{2} & \cos \frac{\theta}{2} & 0 \\
i\sin \frac{\theta}{2} & 0 & 0 & \cos \frac{\theta}{2}
\end{array}\right]
\label{eq:ryy}
\end{equation}
\end{small}

\begin{small}
\begin{equation}
R_{ZZ}(\theta)=\left[\begin{array}{cccc}
e^{-i\frac{\theta}{2}} & 0 & 0 & 0\\
0 & e^{-i\frac{\theta}{2}} & 0 & 0 \\
0 & 0 & e^{-i\frac{\theta}{2}} & 0 \\
0 & 0 & 0 & e^{-i\frac{\theta}{2}}
\end{array}\right]
\label{eq:rzz}
\end{equation}
\end{small}

\subsection{Controlled Gates}


In addition to two-qubit gates that involve rotations, there are also gates that utilize a control qubit and a target qubit. These gates, known as controlled gates, perform specific operations on the target qubit based on the value of the control qubit. We introduce three main types of controlled gates below.

{\noindent \bf CNOT Gate:}
The CNOT gate is a commonly used example of a two-qubit gate in quantum computing. It operates by flipping the value of the target qubit if and only if the control qubit is measured as 1. Otherwise, it leaves the target qubit unchanged. The CNOT gate can be represented by the following matrix:

\begin{equation}
CNOT=\left[\begin{array}{cccc}
1 & 0 & 0 & 0 \\
0 & 0 & 0 & 1 \\
0 & 0 & 1 & 0 \\
0 & 1 & 0 & 0
\end{array}\right].
\label{eq:squ-3}
\end{equation}



{\bf \noindent Controlled Rotation Gates:}
Equations \eqref{eq:crx}--\eqref{eq:crz} represent controlled rotation gates in matrix notation. These gates are similar to the CNOT gate, but instead of flipping the state of the target qubit, they apply a rotation operation based on the measurement of the control qubit. 

\begin{equation} 
\centering
CR_X(\theta)=\left[\begin{array}{cccc} 
1 & 0 & 0 & 0 \\
0 & 1 & 0 & 0 \\
0 & 0 & \cos\frac{\theta}{2} & -\sin\frac{\theta}{2} \\
0 & 0 & -\sin\frac{\theta}{2} & \cos\frac{\theta}{2} 
\end{array}\right]
\label{eq:crx}
\end{equation}

\begin{equation} 
\centering
CR_Y(\theta)=\left[\begin{array}{cccc} 
1 & 0 & 0 & 0 \\
0 & 1 & 0 & 0 \\
0 & 0 & \cos\frac{\theta}{2} & -\sin\frac{\theta}{2} \\
0 & 0 & \sin\frac{\theta}{2} & \cos\frac{\theta}{2} 
\end{array}\right]
\label{eq:cry}
\end{equation}

\begin{equation} 
\centering
CR_Z(\theta)=\left[\begin{array}{cccc} 
1 & 0 & 0 & 0 \\
0 & 1 & 0 & 0 \\
0 & 0 & e^{\frac{i\theta}{2}} & 0 \\
0 & 0 & 0 & e^{\frac{i\theta}{2}} 
\end{array}\right]
\label{eq:crz}
\end{equation}

The CSWAP gate offers an advantage by requiring only the measurement of the ancilla qubit. When qubits are directly measured, their states collapse, resulting in the loss of superposition. However, with the SWAP test and the use of the CSWAP gate, it becomes possible to maintain the superposition of the other qubits. This is achieved by measuring the quantum state fidelity through the ancilla qubit instead of directly measuring the qubits themselves. Consequently, this approach minimizes the information lost during measurement. By leveraging the CSWAP gate and the SWAP test, the superposition and entanglement of the qubits can be preserved, allowing for accurate fidelity measurements without compromising the quantum state.



\subsection{Grover's Algorithm}\label{sec:groversalg}

In classical computing, the linear or sequential search algorithm is typically employed to search through an unsorted database. This approach involves examining each element in the database until the desired item is found. Consequently, the time complexity of this algorithm is $O(N)$, where $N$ represents the number of items in the database. On average, this method requires searching through approximately  $\frac{N}{{2}}$ elements before finding the target item.
In the quantum domian, Lov Grover proposed an algorithm that harnesses the unique properties of quantum computing to achieve a quadratic speed improvement for unstructured search problems. Grover's algorithm exhibits a complexity of $O(\sqrt{N})$.
The algorithm employs a black box function called an oracle to facilitate the search process. This oracle alters the phase of the solution states, effectively allowing the algorithm to home in on the desired solution more efficiently. The oracle is expressed as a diagonal matrix.
In a two-qubit system, if the desired state is $|11\rangle$, then the oracle, $U_{t}$, admits the following form:
\begin{equation}
    U_{t}=
    \begin{bmatrix}
    1&0&0&0\\
    0&1&0&0\\
    0&0&1&0\\
    0&0&0&-1
    \end{bmatrix}.
\end{equation}
For each state $|x\rangle$ in the list, we have
\begin{equation}
    U_{t}|x\rangle=
    \begin{cases}
          |x\rangle, &  \text{for} \quad x\neq t,\\
         -|x\rangle, &  \text{for} \quad x= t,\\
    \end{cases}  
\end{equation}
where $t$ is our target item $|11\rangle$.

The algorithm uses amplitude amplification to increase the probability of measuring the target state. Beginning from a uniform superposition $|s\rangle$, which can be expressed as $|s\rangle=H^{\otimes n}|0\rangle^n=\frac{1}{\sqrt{N}}\sum_{x=0}^{N-1}|x\rangle$, vectors $|s\rangle, |s'\rangle$ and $|t\rangle$, can be used to calculate the reflection angle $\theta$. Assume vectors $|s'\rangle$ and $|t\rangle$ are perpendicular, then $|s\rangle=sin\theta|t\rangle+cos\theta|s'\rangle$. By solving for $\theta$, the reflection angle can be expressed by~\cite{eleanor2011quantum}
\begin{equation}
    \theta=\arcsin\langle s|t\rangle=\arcsin\frac{1}{\sqrt{N}}.
\end{equation}

Next, we use the oracle reflection $U_f$ to reflect the state $|s\rangle$ across the $|s'\rangle$. This switches the amplitude in front of the $|t\rangle$ to negative. Then, we use another reflection $U_s$ on the state $|s\rangle$ where $U_s=2|s\rangle\langle s|-1$ to map the state to $U_sU_f|s\rangle$. These two reflections make the state $|s\rangle$ closer to the state $|t\rangle$ and decrease the amplitudes of the other states. We repeat these two steps $n$ times and obtain the state $|\psi_n \rangle$, where $|\psi_n \rangle=(U_sU_f)^n |s\rangle$~\cite{eleanor2011quantum}.

\begin{figure}[!t]
    \centering
    \includegraphics[width=1\linewidth]{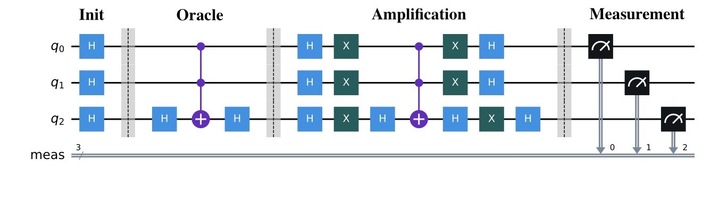} 
    \caption{Grover's Algorithm with 3 qubits} \label{grover circuit_3}
\end{figure}

{\bf  \noindent A 3-Qubit Example:} Figure~\ref{grover circuit_3} represents the quantum circuit with associated gate operations for a 3-qubit Grover's algorithm. 
The qubits required for Grover’s algorithm can be calculated by $n=\log_2N$. In this example, we have $N=8$, which requires $n=3$ qubits.  The reflecting angle $\theta$ is
\begin{equation}
    \theta=\arcsin\frac{1}{\sqrt{8}}.
\end{equation}

Let us start with finding one target state, e.g., $|111\rangle$. As shown on Figure~\ref{grover circuit_3}, Hadamard gates are placed to all the qubits to put them onto the superposition state, yielding the quantum state $|\phi_{1}\rangle$ at this moment, 
\begin{equation}
\begin{split}
    |\phi_{1}\rangle=\frac{1}{\sqrt{8}}(|000\rangle+|001\rangle+|010\rangle+|011\rangle+|100\rangle
    \\
    +|101\rangle+|110\rangle+|111\rangle).
\end{split}
\end{equation}


Then, we use an oracle function to mark the target state, which leads to state $|\phi_{2}\rangle$
\begin{equation}
\begin{split}
    |\phi_{2}\rangle=\frac{1}{\sqrt{8}}(|000\rangle+|001\rangle+|010\rangle
    +|011\rangle+|100\rangle
    \\
    +|101\rangle+|110\rangle-|111\rangle).
\end{split}    
\end{equation}


The third step is to apply the amplitude amplification process to the qubits based on the following procedure:
\begin{enumerate}
    \item Apply the Hadamard gates and the Rotation $X$ gates to all three qubits;
    \item Apply a controlled $Z$ gate with the control qubits $q_0$ and $q_1$ and the target qubit $q_2$;
    \item Apply the Hadamard gates and the Rotation $X$ gates to all three qubits again.
\end{enumerate}

\begin{figure}[!t]
    \centering
    \includegraphics[width=1\linewidth]{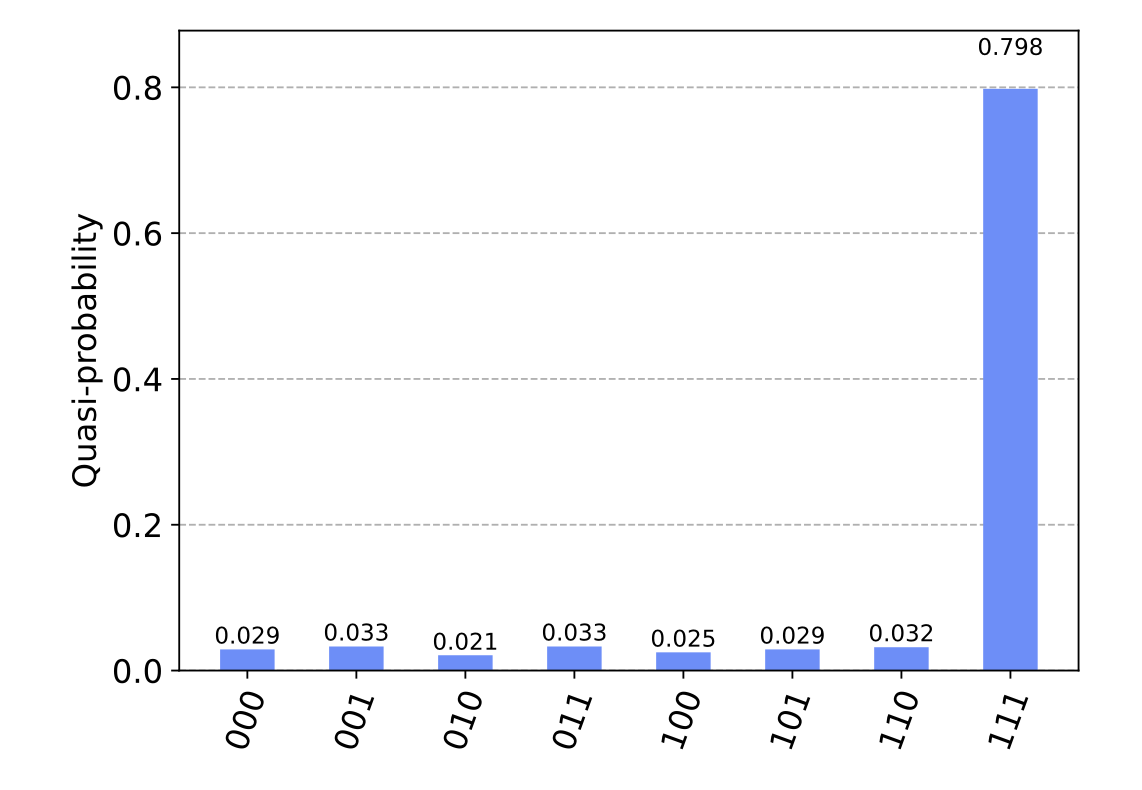} 
    \caption{3-Qubit Example: State Probabilities} \label{grover_p}
\end{figure}

Finally, we get probability distribution of the quantum states
as shown in Figure \ref{grover_p}.
Please note that under a larger dataset, the amplification process will be repeated multiple times to obtain targeted states. 

Grover’s algorithm can be used to find multiple target states in a list by resorting to the transformation of statevector. By initializing the statevector with all states in $|0\rangle$, we can switch the state $|0\rangle$ to state $|1\rangle$ if it is a target state.
For example, if the target state is $|000\rangle, |010\rangle$ and $|111\rangle$, the statevector transformation will be
\begin{equation}
    [0,0,0,0,0,0,0,0] \rightarrow [1,0,1,0,0,0,0,1].
\end{equation}

When the quantum system is measured, we will observe the target state with a higher probability compared to others. 
However, to increase the difference of probabilities between searching targets and others, the amplification process will be repeated multiple times. 
With each repetition, target states' probability continues to increase. 
The number of repetitions required depends on several factors, including the number of target states, the size of the dataset, and the number of qubit. 
\section{\soll~ Solution Design}
This section presents our solution design for quantum search, including the \soll~system architecture and the algorithms. 

\subsection{System Architecture}

We first introduce the problem setting and then describe the system architecture of \soll, a quantum-classical system designed for searching targeted (index, value) pairs. 
The system receives a set of $n$ initial values, ${v_1,...v_i,...v_n} \in D$, each of which is associated with an index to form the (index, value) pairs. The set of targets, ${v_i,...v_j} \in T$, is specified by the users, and the objective is to find the corresponding indexes of the targets.

To achieve this goal, we utilize Grover's search algorithm, which is described in Section \ref{sec:groversalg}. 
However, the original algorithm cannot be directly applied. First, it only amplifies the amplitude of the targeted states and does not output the targets and indexes. Moreover, since both the indexes and values need to be encoded, the qubit requirement is higher than that of value-only searches. Therefore, our goal is to output the targeted (index, value) pairs while reducing the number of required qubits.

We propose a hybrid approach that combines quantum and classical computing resources. Specifically, the system consists of a quantum part and a classical part. The quantum part is responsible for implementing Grover's search algorithm to amplify the amplitude of the targeted states, and the classical part performs postprocessing to extract the (index, value) pairs and reduces the required number of qubits.

\begin{figure}[!t]
\centering
\includegraphics[width=1\linewidth]{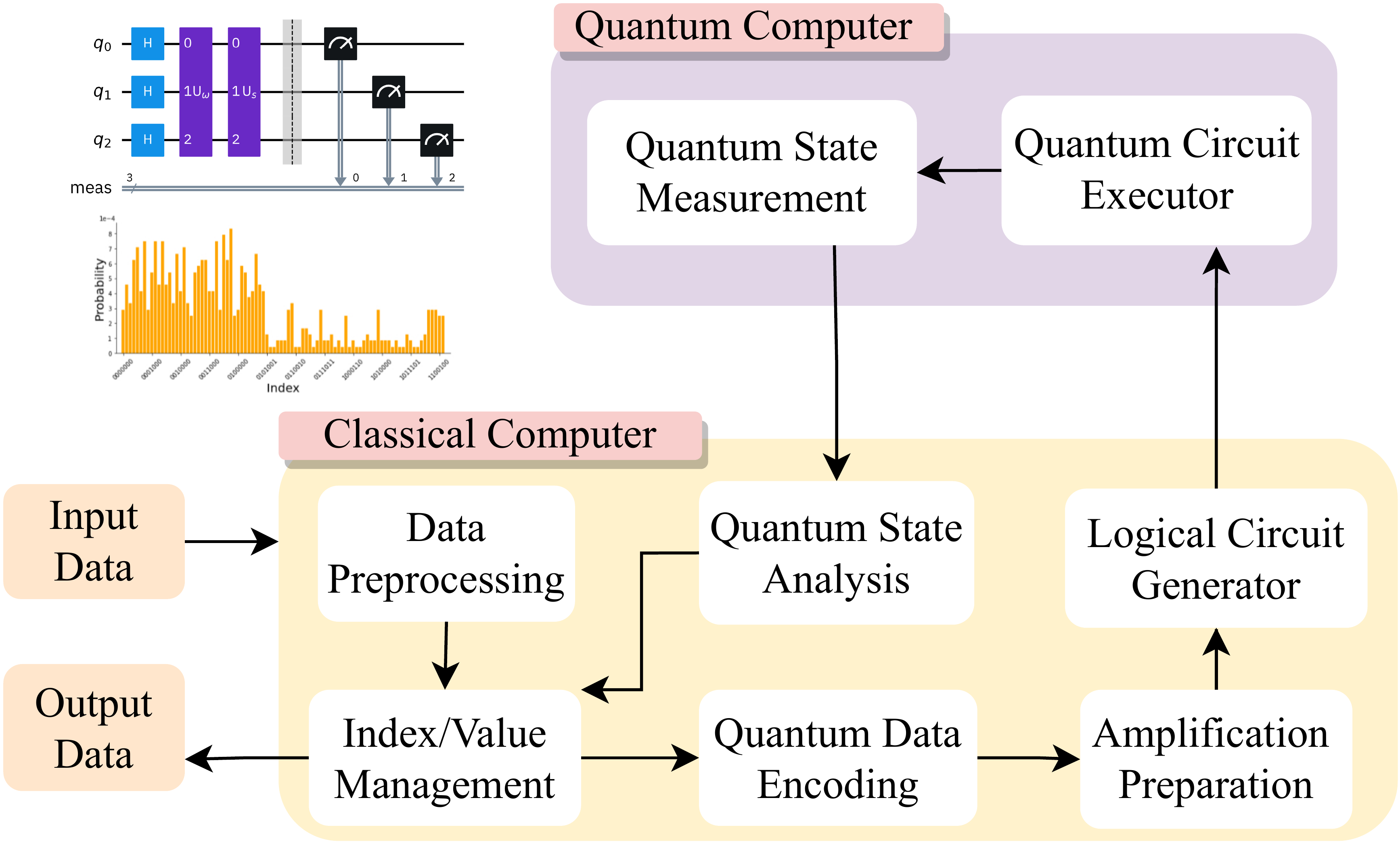}
\caption{\soll~System Architecture}
\label{fig:system}
\end{figure}

The hierarchical structure of \soll's system architecture is displayed in Figure \ref{fig:system}, which demonstrates a quantum-classical system design. Input data, e.g., unstructured databases, is submitted to the classical component to the \textbf{Data Preprocessing} module. During this phase, the module analyzes the data's size, index ranges, and extracts metadata for subsequent processing.

The \textbf{Index/Value Management} module receives the data
from two possible sources, (1) Data Preprocessing and (2) Quantum State Analysis. Its functionalities divided into two scenarios. When the data comes from Data Preprocessing module, the system employs its original index and value associations. Otherwise, the module regenerates new indexes and values after filtering out unlikely data points, as informed by the Quantum State Analyst module. 
In this case, the module maintains a mapping between generated and original indexes/values in order to return the original values at the end. Algorithm~\ref{alg:UI} Updating Indexes, Algorithm~\ref{alg:UV} Updating Values and Algorithm~\ref{alg:MI} maintain the original Indexes are implemented in this module which will be described in detail in the next subsection.

The \textbf{Quantum Data Encoder} module accepts the processed data from the Index/Value Manager and encodes the classical data to quantum qubits. As our data updates every iteration, the encoding should be updated accordingly. The \textbf{Amplification Preparation} module calculates the number of amplitude amplifications, e.g., reflections, based on the iteration number and previous results. This number is used by the \textbf{Logical Circuit Generator} to prepare a logical circuit, e.g, construct the oracle function, for quantum search.

The generated logical circuit is passed to a quantum computer, where Grover's search algorithm is conducted with a given number of amplitude amplification within the \textbf{Quantum Circuit Executor} module. The quantum states are measured by the \textbf{Quantum State Measurement} module and transferred back to the classical component for further processing at the end of the search.

Upon receiving the results, the {\bf Quantum State Analyst} module is activated to perform Algorithm~\ref{alg:F} Filter. If the algorithm finds all solutions, their original indexes will be returned. Otherwise, it conducts filtering to ensure that only potential solutions enter the next iteration. With this feature, the problem set is reduced when the Index/Value Management modules are called in the second and following rounds. Consequently, fewer qubits are required to continue the search. Meanwhile, the mappings between current and original indexes and values are maintained.

\subsection{Algorithms Design}
In this subsection, we describe the details of our algorithms in the system. Table \ref{table:notation} summarizes the notations adopted in the algorithms.
\soll~ aims to reduce the problem size iteratively, which leads to a reduction of required qubits. 
The system achieves the reduction by updating the indexes and values of the input data at each iteration, except for the first one that is input by clients. 
At the end, however, it returns the original (index, value) pairs. Therefore, \soll~ has to preserve the mapping between original (index, value) pairs with generated ones. 

\begin{table}[ht]
	\centering
	\caption{Notation Table}
	\scalebox{0.75}{
		\begin{tabular}{ | c | c |  }				
			\hline
            $k$ & the updating indexes of the original index
            \\
            \hline
            $PM$ & The dictionary map data set in the last iteration. \\
            \hline
            $Q_j$ & The qubits that we need at iteartion $j$. \\
            \hline
			$d_i$ & The $i^{th}$ element in the input data set. \\
			\hline
			$D_j$ & The input data set at iteration $j$.\\
			\hline
            $p_i$ & The $i^{th}$ element in the binary number set. \\
            \hline
			$P_j$ & The [index, value] pair binary number set which is generated by $D_j$ at iteration $j$. \\
            \hline
            $m_i$ & The $i^{th}$ element in the dictionary map. \\
            \hline
            $M_j$ & The dictionary map that pairs the index and value at iteration $j$. \\
            \hline
            $GBP(d)$ & The function that generates [index, value] pair \\
            &binary number set by input data set $d$. \\
            \hline
            $GM(d)$ & The function that generates the dictionary map \\
            &that pairs the index and value by input binary data set $d$. \\
            \hline
            $GI(a,B)$ & The function that gets the index of input $a$ in a list $B$. \\
            \hline	
            $H(a)$ & The function that generates the hash value of the input $a$.\\
            \hline
            $MapI$ & The dictionary map that pairs the current indexes $k$ and original indexes $m_i$.\\
            \hline
            $GP_j$ & The probability dictionary that is generated by Grover's Algorithm at iteration $j$.\\
            \hline
            $gp_i$ & The $i^{th}$ element in probability dictionary. \\
            \hline
            $CD(a,b)$ & The function that computes the distance between $a$ and $b$. \\
            \hline
            $CQ(d)$ & The function that computes the required qubits by input data set $d$.\\
            \hline
            $nce_i$ & the mean value of $c_i$
            \\
            \hline
            
		\end{tabular}	
	}	
	\label{table:notation}
\end{table}

\begin{algorithm}[!t]
\caption{Updating Indexes, UI($M_j$)}
\begin{algorithmic}[1]
\STATE  Input: $M_j$
\STATE  Initialization: $k=0$
\FORALL{$m_i \in M_j$}
    \STATE $MapI \gets [m_i(i), k]$
    \STATE $k++$
\ENDFOR
\RETURN $MapI$
\end{algorithmic}
\label{alg:UI}
\end{algorithm}
\subsubsection{Index Updates and Mapping Maintenance} 
Algorithm 1 is used to generate new indexes for input data $M_j$, which is entered by client, at the beginning of each iteration except for the first one. 
It first accepts the input data $M_j$ and initialize the index $k$ (Line 1-2).
Then, the algorithms creates a new dictionary $MapI$ to store the current index $k$ from $0$ and the original index $m_i$ in dictionary to maintain the mapping between them (Line 3-6).
Lastly, it return the new dictionary $MapI$ that will be used by Algorithm 3 (Line 7).
\begin{algorithm}[!t]
\caption{Updating Values, UV($D, T$)}
\begin{algorithmic}[1]
\STATE  Inputs: $D$: Input dataset, $T$: Searching targets
\STATE  Initialization: $L = [~], C = [~]$
\STATE  $C \gets H(T)$
\FORALL{$d_i \in D$}
    \IF{$H(d_i) \in C$}
        \STATE $L \gets GI(H(d_i), C)$
    \ELSE
        \STATE $C \gets H(d_i)$
        \STATE $L \gets GI(H(d_i), C)$
    \ENDIF
\ENDFOR
\RETURN $L$

\end{algorithmic}
\label{alg:UV}
\end{algorithm}

\subsubsection{Value Updates} Algorithm~\ref{alg:UV} aims to update remaining values after each iteration so that the searching range in the input dataset can be reduced to the number of different values. 
It first accepts input data set $D$, and the target value $T$ (Line 1).
The system initializes two empty lists, $L$ and $C$.
and appends the hash value of the target to $C$ (Line 2 -3).
Then, \soll~ directs into two branches for each date point in $D$ based on the hash value. (1) If the hash value is in $C$, its index in $C$ is appended to $L$ as its new value; (2) If its hash value is not in $C$, its hash value is appended to $C$ first, then its value is updated to the index of its hash value in $C$, and appended to $L$ (Line 4-11). Finally, it returns the output list $L$ of updated values (Line 12).

\begin{algorithm}[!t]
\caption{Maintain Indexes, MI($M_j, MapI$)}
\begin{algorithmic}[1]
\STATE  Inputs: $M_j, MapI$
\STATE  Initialization: $M=\{~\}$
\FORALL{$m_i \in M_j$}
    \STATE $M \gets [MapI, m_i(value)]$
\ENDFOR
\RETURN $M$
\end{algorithmic}
\label{alg:MI}
\end{algorithm}
\subsubsection{Maintain Index Mapping}
While Algorithm 1-2 reduce the searching range by reorganizing the indexes and values. However, \soll~ is designed to return the original (index, value) pairs to the clients. Therefore, the system requires to dynamically manage a mapping between generated (index, value) pairs with the original ones. 

Algorithm~\ref{alg:MI} maintains the mapping in each iteration except for the first one.
Initially, it accepts the dictionary map $M_j$ that pairs the index and value at iteration $j$, and the dictionary map $MapI$ that pairs the current indexes $k$ and original indexes $m_i$. These information come from the previous iteration (Line 1). 
Then, it initializes the empty dictionary $M$ (Line 2).
\soll~ stores each original index which corresponds to the element $m_i$ from $MapI$ to the new dictionary $M$ (Line 3-5).
It returns the generated dictionary with updated mappings, $M$ (Line 6).

\begin{algorithm}[!t]
\caption{Filter, F($GP_j$)}
\begin{algorithmic}[1]
\STATE  Inputs: $GP_j$
\STATE $c1 = [~], c2 = [~]$
\STATE $ce1 = gp_1$
\STATE $ce2 = gp_n$
\REPEAT
\FORALL{$gp_i \in GP_j$}
    \IF{$CD(gp_i, ce1) < CD(gp_i, ce2)$}
        \STATE $c1 \gets gp_i$
    \ELSE
        \STATE $c2 \gets gp_i$
    \ENDIF
\ENDFOR
\STATE $nce1 = sum(c1)/len(c1)$
\STATE $nce2 = sum(c2)/len(c2)$
\UNTIL{{$ce = nce~\OR~i = mi$}}
\IF{$ce1 > ce2$}
    \RETURN $c1$
\ELSE
    \RETURN $c2$
\ENDIF
\end{algorithmic}
\label{alg:F}
\end{algorithm}
\subsubsection{Filtering Data Points}
The Algorithms 1-3 focus on updating (index, value) pairs and maintaining the mappings. While these techniques can reduce the range of search space, they fail to reduce the data points within the dataset. To further reduce the problem size, we propose a self-adaptive filtering algorithm to process the state probabilities after each quantum search. 

Algorithm~\ref{alg:F} presents our filtering algorithm. Specifically, it categorizes the data points into two clusters, potential targets and non-targets.
The input data is the state-probability dictionary $GP_j$ that is generated by quantum search  at iteration $j$ (Line 1). It then initializes two clusters $c1$ and $c2$ (Line 2).
The maximum and minimum values in the probability dictionary are used as the centroid of the two clusters (Line 3-4).
Lines 5-15 represent a processing in each iteration. 
The system first calculates the distance between each value in the probability dictionary and the two centroids, and assign them to the closer cluster. After the assignments, we calculate the mean of the clusters to obtain new centroids. If the new centroids are the same as the original centroids, it means that we have reached convergence, and the loop ends. We also set a maximum number of loops to prevent infinite looping.
Eventually, it return the cluster that has a larger centroid (Line 16-20).
The principle of this algorithm is that in the probability dictionary, the probability of the targets is much larger than that of the non-targets.

\subsubsection{Integrated Iterative Quantum Search}
Based on the Algorithms 1-4, the entire quantum-classical search process in \soll~ is shown in Algorithm~\ref{alg:RESAQuS}.
Initially, the input data set $D$, target values $T$, and maximum iteration number $J$ are loaded (Line 1).
The first iteration is different from the second and following iterations as it processes the original (index, value) pairs.
It first calls Algorithm~\ref{alg:UV} to update the values of the input data $D$.
However, since Algorithm~\ref{alg:UV} defaults the target values from $0$ to $n-1$, which the $n$ is the number of the searching targets, these updated target value corresponds to Algorithm~\ref{alg:UI}.
Then, $CQ$ calculates the number of qubits required for this iteration $Q_1$ (Line 4). Next, $GBP$ converts the input data $D_1$ to binary data $P_1$ (Line 5). After that, we run Grover's search with input binary data $P_1$ (Line 6). Afterward, the system obtains the generated probability dictionary $GP_1$ using Algorithm~\ref{alg:F} and converts the probability dictionary $GP_1$ into a dictionary of indexes and values $M_1$  (Line 7-8). At the end of the first iteration, we update the maximum iteration number $j$ to $1$, and initialize the dictionary map data set $PM$ which will be in previous iteration (Line 9-10).

From the second iteration onwards (Line 11-23), we start the while loop with conditions. After each iteration, if the results of this iteration differ from those of the previous one and the maximum number of iterations has not been reached, the next iteration will be performed (Line 11). If the results of this iteration are the same as those of the previous iteration or the maximum number of iterations has been reached, the iteration stops. Then, it stores the results of the previous iteration and uses Algorithm~\ref{alg:UI} to store the original indexes of the input data in the beginning of the iteration (Line 12-13). Next, it updates the values of the input data $D_j$ using Algorithm~\ref{alg:UV} (Line 14-15). After we have the updated input $D_j$, we did the same steps as the first iteration. It calculates the number of qubits required for this iteration $Q_j$ with $D_j$, converts the input data $D_j$ to binary data $P_j$, runs Grover's algorithm with input binary data $P_j$, filters the generated probability dictionary $GP_j$ using Algorithm~\ref{alg:F}, and converts the probability dictionary $GP_j$ into a data dictionary of indexes and values $M_j$ (Line 16-21). Finally, It maps the original indexes of the input data before the end of the iteration by using Algorithm~\ref{alg:MI} (Line 21). At the end of the loop, it also update the iteration number $j$ by adding $1$ (Line 22). Ultimately, it returns the targeted (index, value) pairs (Line 24).

\begin{algorithm}[!t]
\caption{Iterative Quantum Search, IQS($D, T, J$)}
\begin{algorithmic}[1]
\STATE  Input: 
\item[] $D:$ input dataset
\item[] $T:$ searching targets
\item[] $J:$ maximum iteration number
\STATE $D_1 = UV(D, T)$
\STATE $T = [0,...n-1]$
\STATE $Q_1 = CQ(D_1)$
\STATE $P_1 = GBP(D_1)$
\STATE $GP_1 = QuantumSearch(P_1)$
\STATE $GP_1 = F(GP_1)$
\STATE $M_1 = GM(GP_1)$
\STATE $j = 1$
\STATE $PM = \{~\}$
\WHILE{$PM \neq M_j \And j < J$}
    \STATE $PM = M_j$
    \STATE $MapI = SI(M_j)$
    \STATE $D_j = M_j(value)$
    \STATE $D_j = UV(D_j, T)$
    \STATE $Q_j = CQ(D_j)$
    \STATE $P_j = GBP(D_j)$
    \STATE $GP_j = Grover'sSearch(P_j)$
    \STATE $GP_j = F(GP_j)$
    \STATE $M_j = GM(GP_j)$
    \STATE $M_j = RI(M_j, MapI)$
    \STATE $j++$
\ENDWHILE
\RETURN $M_j$
\end{algorithmic}
\label{alg:RESAQuS}
\end{algorithm}
\section{Evaluation}


In this section, we evaluate \soll~ with different settings and configurations. 


\subsection{System Implementation and Settings}

\soll~ is implemented in Python 3.8 and the IBM Qiskit Quantum Computing simulator package. We use the Aer simulator as our backend to simulate a {\em noise-free} environment. Grover's algorithm is implemented using Qiskit's amplitude amplifiers APIs~\cite{apidoc}, and the number of shots is set to 24000.

As mentioned in Section IV, we convert the input data into binary values to run Grover's algorithm. Each binary value consists of an (index, value) pair: the first part is its index and the second part is its value. For example, $001001$ has 4 index qubits and 2 value qubits, indicating that its index is 2 with a value of 1. The evaluation considers two types of systems,
\begin{itemize}
    \item {\bf Standalone Mode}: one classical machine combined with one quantum machine in the system. In this configuration, the client, represented by the classical machine, communicates with the quantum machine directly.
    
    \item {\bf Cluster Mode}: multiple classical machines integrate with multiple quantum machines. In this configuration, the clients, represented by multiple classical machines, submit searching jobs to a classical manager, who further distributes the workload to multiple quantum workers. 
\end{itemize}

\subsection{Evaluation Metrics}

In the experiments, we compare \soll~ with two state-of-the-art solutions in the literature. 
\begin{itemize}
    \item {\bf GSearch}~\cite{qiskit-Grover}: The Grover's Search with IBM Qiskit implementation. It is a quantum-only solution that utilizes a single iteration with the optimal number of invocations of Grover's operator, e.g., amplitude amplification process. 

\item {\bf IQuCS}~\cite{mu2022iterative}: The Grover's Search with iterative post-processing on classical computers. It is a threshold-based solution, such that a pre-defined threshold is employed to filter out non-target in each iteration. 

\end{itemize} 

We conduct the same experiments with {\bf GSearch, IQuCS} and {\bf \soll} to compare their performance.  Specifically, $optimal\_num\_iterations$ method in the Grover APIs is used to determine the optimal number of iterations for Grover's algorithm~\cite{optimal}. The both of GSearch and  IQuCS algorithms are threshold based.  GSearch's threshold set $0.01$ to ensure finding the target by filtering the non-target, and {\bf IQuCS}'s thresholds are 1 times of mean value of probability. There will be a special case of  IQuCS's threshold, which will be explained in the following subsection. 

Furthermore, we utilize three key metrics to analyze and compare the results: (1) accuracy: the percentage of targets that have been successfully identified; (2) total number of invocations: the number of invocations of Grover’s operator, which is repeated amplitude amplification in Grover’s algorithm, and (3) cumulative qubits consumption (CQC): the value that represents the qubit consumption of an iterative quantum-classical search algorithm. 
Specifically, the $CQC$ is defined by
\begin{equation}
CQC = \sum_{i=1}^{i=n} C_i \times N_{q_i},
\end{equation}
where $i$ is the iteration number, $C_i$ is the number of Grover's operator invocations at iteration $i$, and $N_{q_i}$ is the number of qubits at iteration $i$. A larger value of CQC indicates a more resource-intensive algorithm. 
In GSearch,  $CQC = Q \times I$, where $Q$ is the qubits required by the algorithm and $I$ is the optimal number of iterations. In \sol~and \soll, $CQC = \sum_{i=1}^{i=n} I_i \times Q_i$, where $I_i$ is the number of Grover's algorithm invocation at iteration $i$ and $Q_i$ is the number of qubits required at iteration $i$. In \sol, the invocation number equals 1 when the iteration number is odd. Otherwise, the invocation number equals 2. In \soll, the invocation number always equals 1.


\begin{figure*}[ht]
\centering
\includegraphics[width=\linewidth]{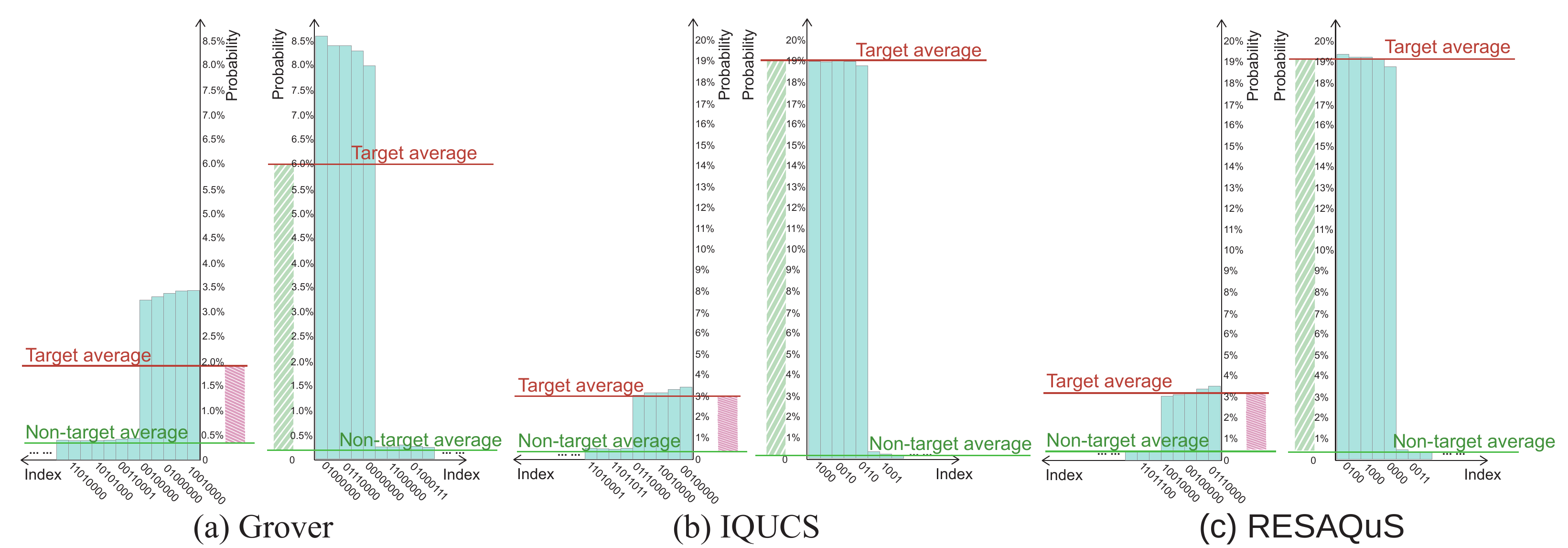}
\caption{15-5 Experiments: (a) Probabilities of GSearch in Iteration-1-Invocation-1 (Left) and Iteration-2-Invocation-2 (Right); (b) Probabilities of \sol~ in Iteration-1-Invocation-1 (Left) and Iteration-3-Invocation-2 (Right); (c) Probabilities of \soll~ in Iteration-1-Invocation-1 (Left) and Iteration-2-Invocation-2}
\label{e1}
\end{figure*}

\subsection{Standalone Mode}
In this setting, we have one classical machine and one quantum machine in the system. We conducted three experiments with different data sizes and searching targets.

Each of the three experiments included a graph comparing the results of three algorithms using the same invocation. The X-axis represents the index value, while the Y-axis represents probability, depicted through a blue bar chart. Within the graph, the red line signifies the average probability of the target, the green line represents the average probability of the non-target, and the disparity between them is shaded in red on the left and green on the right.

\subsubsection{ Dataset-15 with 5 searching targets}
In this experiment, our input data size is 15 with 5 searching targets. The initial qubit requirement (e.g., the first iteration) is 8, with 4 qubits for indexes and 4 qubits for values. Figure~\ref{e1}(a) shows the probability distributions of GSearch with different numbers invocations. 
Since GSearch is a single-iteration algorithm, this figure illustrates two individual experiments. 

The left graph of Figure~\ref{e1}(a) displays the results with only 1 invocation. Although the five largest probabilities in the Figure~\ref{e1}(a) are clearly found, GSearch finds $9$ targets in the $1$ iteration, with the largest probabilities are $3.48\%$ and the smallest is $0.42\%$. The average probability of these 9 target states is 2.00\%, while the average probability value of the rest states is 0.33\%, the highest probability value in the non-target is only 0.41\%. 
The right graph displays the distribution with 2 invocations, it is as same as the left. $7$ targets are selected by GSearch this iteration with the largest probabilities are $8.77\%$ and the smallest is $0.31\%$.
The average probability value of the target is 6.09\%, while the average probability value of the non-target is 0.23\%.

As the invocation increases, the difference between the average probability of the target and non-target also increases. It continues to grow until the invocation reaches 5, at which point GSearch can successfully find all 5 targets, resulting in an accuracy of 100\%, and CQC is $8 \times 5$ = 40. By comparing with the other two algorithms, we can see that the target average we finally found is much higher than the non-standard average in Figure~\ref{e1}.

Figure~\ref{e1}(b) shows the results of \sol. As \sol~ is a threshold-based solution, we set the threshold to 1. In the first iteration, 10 states are filtered out, and in the second iteration, no data is filtered out. According to its algorithm, \sol~ reaches the termination condition. The left graph of Figure~\ref{e1}(b) represents the results after the first iteration with 1 invocation. At this point, the average probability of the targets is 3.22\%, and the average probability of the non-targets is 0.33\%. The right graph displays the results after the second iteration, which is when \sol~terminates. The average probability of the targets increases to 19.18\%, and the average probability of the non-targets is 0.37\%. The significant difference between the two clearly indicates the targets. Therefore, the accuracy is 100.00\% in 3 invocations with 2 iterations, and the CQC is $8 \times 1 + 4 \times 2$ = 16.

Figure~\ref{e1}(c) shows the graph of \soll. It also filters out 10 states in the first iteration, which results in 4 qubits in the second iteration. \soll~also finishes the search in 2 iterations and finds 5 target states, with the probability of the targets and the average value the same as the \sol’s. The CQC is $8 + 4$ = 12, which is a 70.00\% reduction of GSearch and a 25.00\% reduction of \sol.

\subsubsection{\bf Dataset-40 with 15 searching targets}
In this experiment, we increase the input data size to 40, with 15 targets to be found. The increased data size requires more qubits. Our initial qubits are 10, with 6 qubits for index and 4 qubits for value.

\begin{figure*}[ht]
\centering
\includegraphics[width=\linewidth]{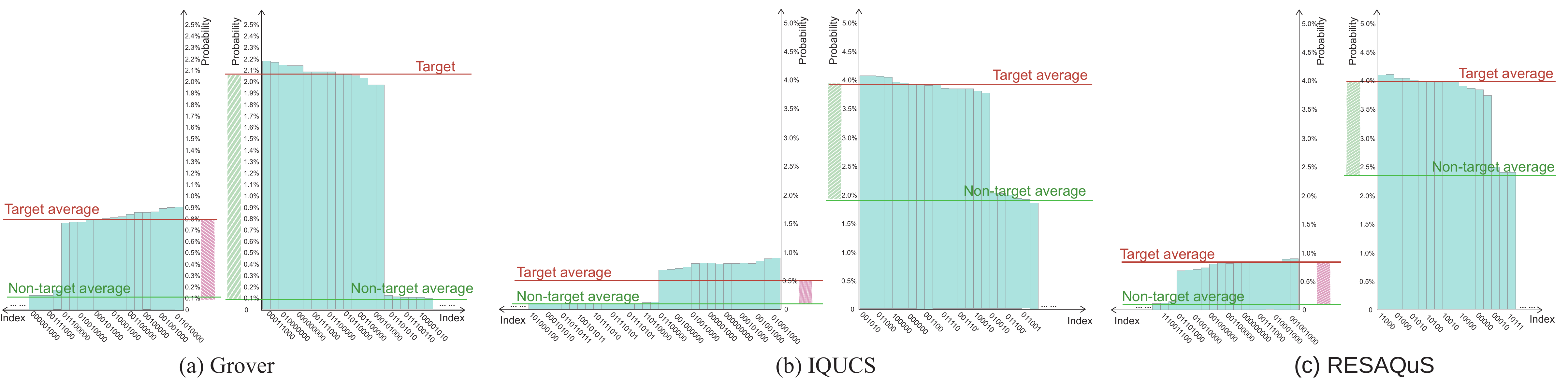}
\caption{40-15 Experiments: (a) Probabilities of GSearch in Iteration-1-Invocation-1 (Left) and Iteration-2-Invocation-2 (Right); (b) Probabilities of \sol~ in Iteration-1-Invocation-1 (Left) and Iteration-3-Invocation-2 (Right); (c) Probabilities of \soll~ in Iteration-1-Invocation-1 (Left) and Iteration-2-Invocation-2}
\label{e2}
\end{figure*}

In Figure~\ref{e2}(a),  the left graph shows the first invocation, in which we find 17 target states at this moment. Their average probability of 0.75\% is much higher than the average non-target states probability of 0.09\%. The right graph shows the second invocation finding 16 targets with an average probability of targets 2.03\%, and an average probability of non-targets 0.07\%. This time, we find 15 targets when the optimal iteration number become 6. The CQC of GSearch is $10 \times 6$ = 60. 

For \sol,  when using 1 as the threshold, we encounter an algorithm error. In the first iteration, 16 data items are filtered out, and 24 remaining data items are sent to the next iteration. In the second iteration, the system filters out 9 data items. It looks like the algorithm find all the targets. However, in the third iteration, 15 target data are filtered out by the system. Since we do not have any data remaining, the system terminates with an error message. The reason for this situation is due to the erroneous threshold selection. The threshold of \sol~is too sensitive, and it is difficult for us to control the threshold in iteration. This fact motivates us to develop \soll~in replace of \sol.

When setting the threshold to 0.8, it uses 4 iterations to find 15 target states. The remaining data after each iteration is 30, 21, 15, and 15. Figure~\ref{e2}(b) shows the results. The target average probability and non-target average probability change from 0.49\% to 3.92\%, and from 0.09\% to 2.42\%, respectively. The CQC is $10+10 \times 2+9+5 \times 2$ = 49  

Figure~\ref{e2}(c) shows the results of \soll. It only takes two iterations to find 15 target values. The average probability of the targets and the average probability of the non-targets are similar to the results of \sol. The accuracy is 100.00\%, and the CQC is $10 + 5$ = 15. This represents a 75.0\% reduction compared to Grover's algorithm and a 69.39\% reduction compared to \sol.

\subsubsection{\bf Dataset-80 with 20 searching targets}
This experiment involves finding 20 targets among 80 data points. Our initial qubits are 12, with 7 qubits for index and 5 qubits for value.

\begin{figure*}[ht]
\centering
\includegraphics[width=\linewidth]{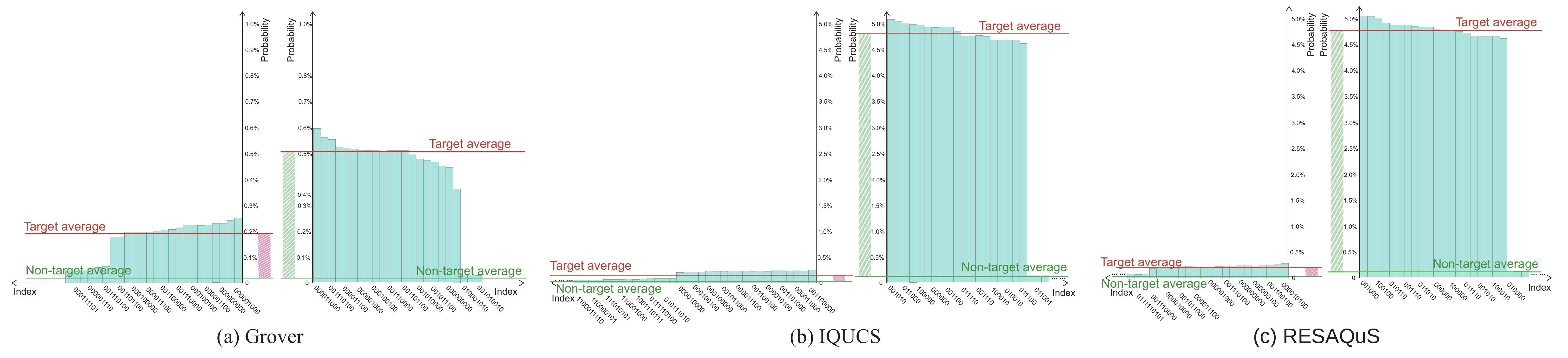}
\caption{80-20 Experiments: (a) Probabilities of GSearch in Iteration-1-Invocation-1 (Left) and Iteration-1-Invocation-2 (Right); (b) Probabilities of \sol~ in Iteration-2-Invocation-1 (Left) and Iteration-3-Invocation-2 (Right); (c) Probabilities of \soll~ in Iteration-1-Invocation-1 (Left) and Iteration-2-Invocation-2}
\label{e3}
\end{figure*}

Figure~\ref{e3}(a) displays the probability distribution graph of GSearch. This time, the final optimal iteration number is 11. We can observe that the optimal iteration number also increases as the number of input data and required qubits increases. The CQC is $12 \times 11$ = 132.  In invocation 1, the system only finds 23 target states. In invocation 2, the system still finds 22 target states. The average target probability increases from $0.20$\% to $0.53$\%, and the average non-target probability decreases from $0.023$\% to $0.021$\%.

Figure~\ref{e3}(b) gives us the process graph of \sol. In this case, we set the threshold to 1. It uses 4 iterations and finds exactly 20 targets. After each iteration, the number of remaining data is 51, 34, 20, and 20. In each iteration, the number of qubits is 12, 10, 8, and 6, respectively. The CQC is $12+10 \times 2+8+6 \times 2$ = 52.

The iterative process of \soll~is Figure~\ref{e3}(c). It still only uses two iterations and finds all 20 targets. In the first iteration, 60 data are filtered out. In the second iteration, the system only uses 6 qubits to implement the algorithm. The result of iteration 2 is the same as the result of iteration 1, so we stop the iteration and find the targets, with an average target probability of 4.79\% and an average non-target probability of 0.09\%. The CQC is $12+6$ = 18, which indicates an 86.36\% reduction of GSearch and a 65.38\% reduction of \sol.

According to three experiments conducted using three different algorithms, /sol employed fewer invocations than GSearch. Not only did /sol use fewer invocations, but it also had fewer iterations. Furthermore, from small to large datasets, the average target probability of /sol and /soll exceeded the values of non-average target probability. From Figure~\ref{e3}(c) (left), we can observe that the difference between the average target probability and non-average target probability for /soll is even greater than that of /sol at the same iteration. Additionally, the average target probability of /soll is more than 2 times greater than the non-average target probability in all experiments, even 10 times greater in Dataset-80.

\subsubsection{\bf Cumulative Qubits Consumption Comparison}
\begin{figure}[htbp]
\centering
    \includegraphics[width=1\linewidth]{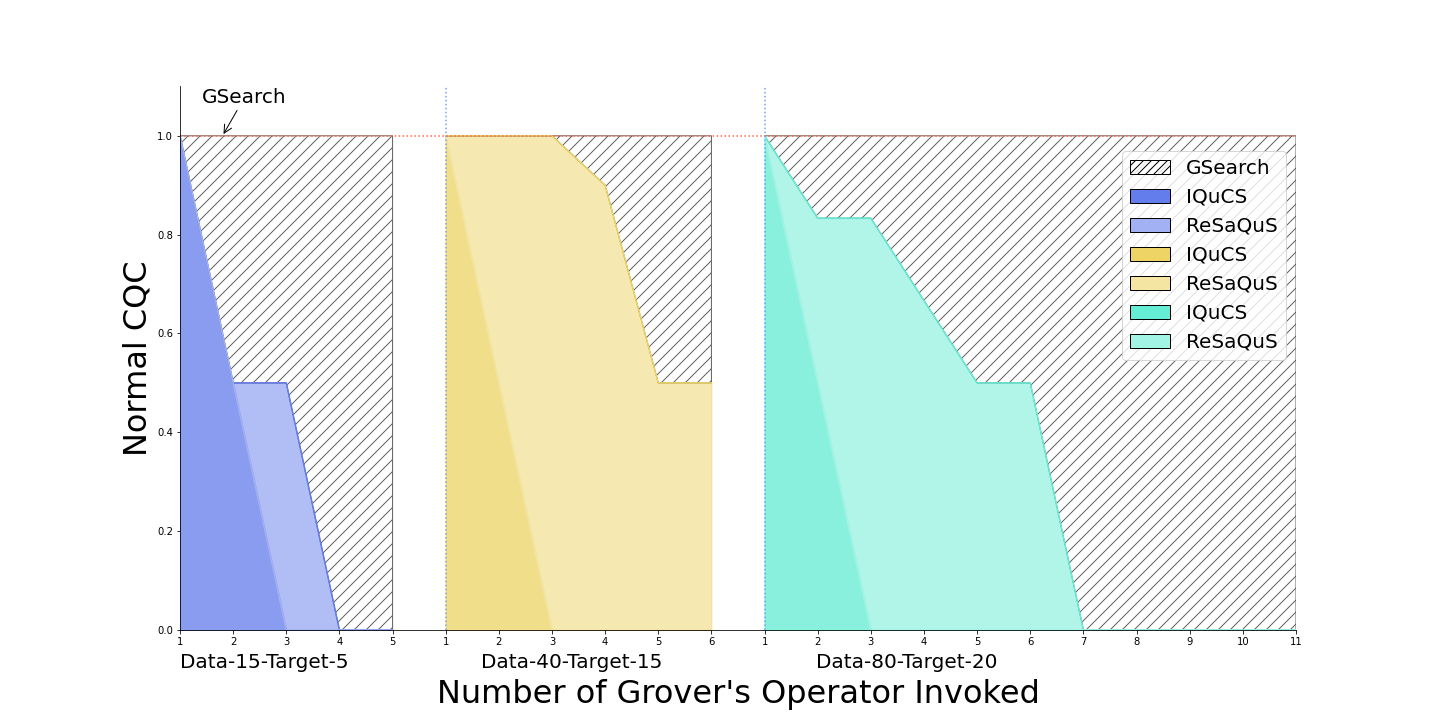}
\caption{Qubits Consumption}
\label{QC}
\end{figure}

Figure~\ref{QC} presents the qubits consumption for GSearch, \sol, and \soll~ in the above mentioned experiments, and the result is normalized with respect to GSearch for comparison. Without post-processing, GSearch utilizes the same number of qubits throughout the entire algorithm. Therefore, its qubits consumption graph is always a rectangle, represented by the shaded area. 
The light area stacked with the dark part is the qubits consumption graph of \sol. 
We can see that it starts reducing the qubits requirement in the second iteration due to post-processing on the classical side. 
The dark area is the qubits consumption graph of \soll. It sharply reduces the required qubits as well as the number of iterations in all experiments. The area presents the CQC value in the figure. Compared to GSearch, the CQC value of \sol~ decreases by 60.00\% and the CQC value of \soll~ decreases by 70.00\% in the first experiment; the CQC value of \sol~ decreases by 18.33\% and the CQC value of \soll~ decreases by 75.00\% in the second experiment; and the CQC value of \sol~ decreases by 60.61\% and the CQC value of \soll~ decreases by 86.36\% in the third experiment. As the data size increases, we can observe that our algorithm \soll~ performs increasingly better than \sol~and GSearch, with increasing savings in the number of iterations, qubits required, and CQC.

\subsection{Cluster Mode}
In this configuration, our hybrid system contains multiple classical machines and multiple quantum machines that form a computing cluster. 

\begin{figure*}[htbp]
\centering
         \begin{subfigure}[b]{0.32\textwidth}
            \centering
                \includegraphics[width=1\linewidth]
                {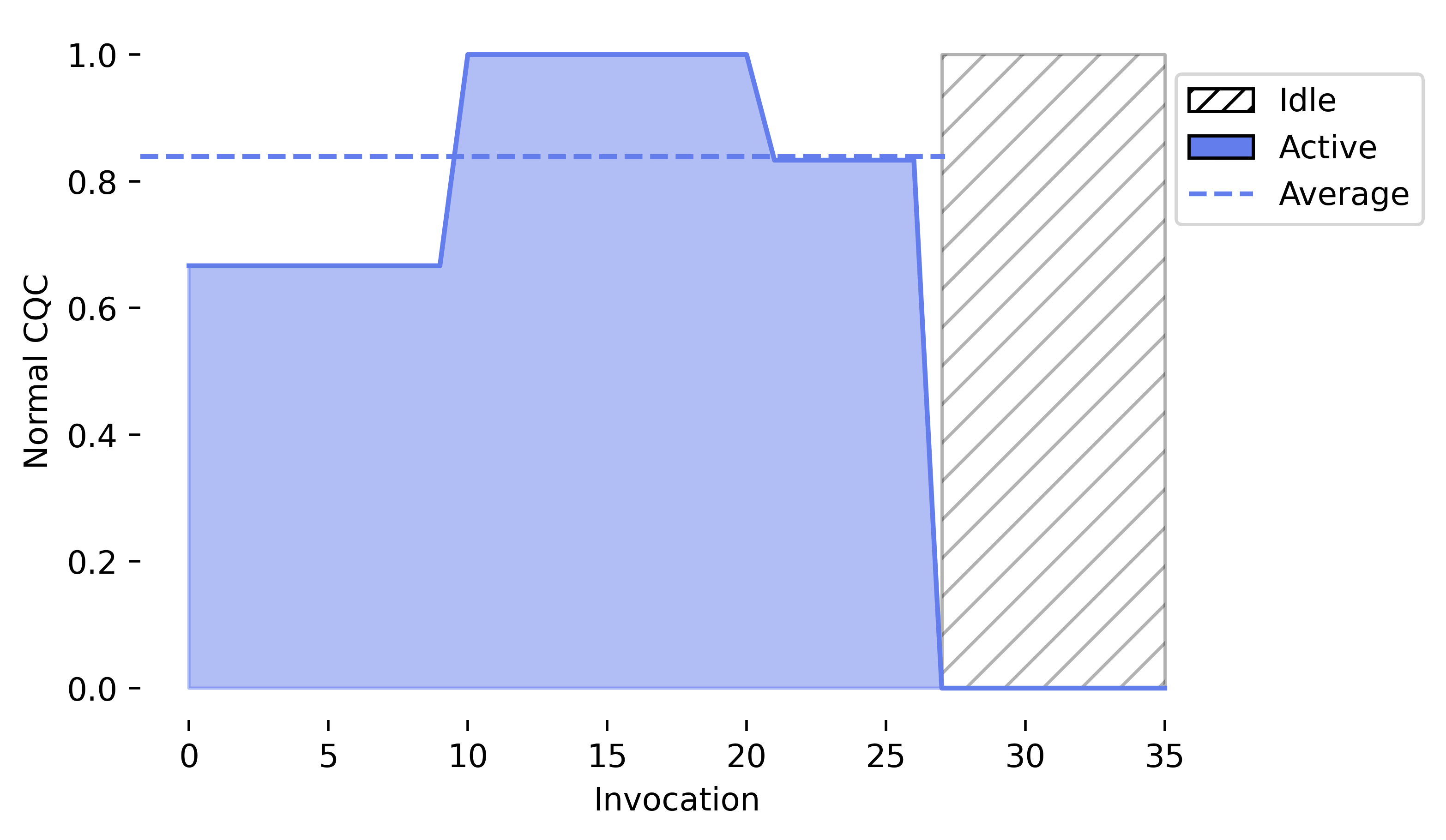}
            \caption{Worker 1}
            \label{f1_w1}
            \end{subfigure}
         \begin{subfigure}[b]{0.32\textwidth}
            \centering
                \includegraphics[width=1\linewidth]
                {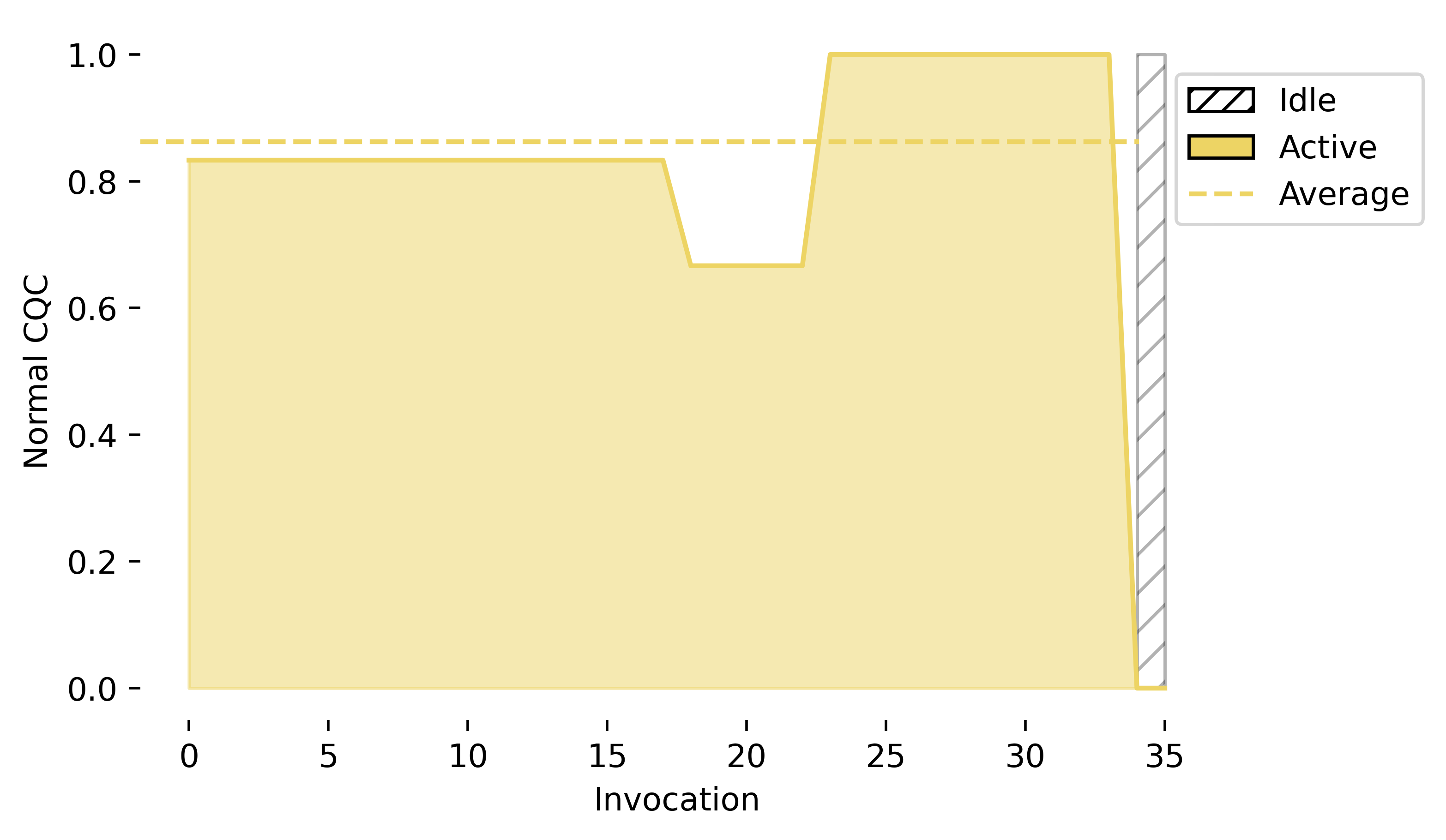}
            \caption{Worker 2}
            \label{f1_w2}
            \end{subfigure}            
         \begin{subfigure}[b]{0.32\textwidth}
            \centering
                \includegraphics[width=1\linewidth]
                {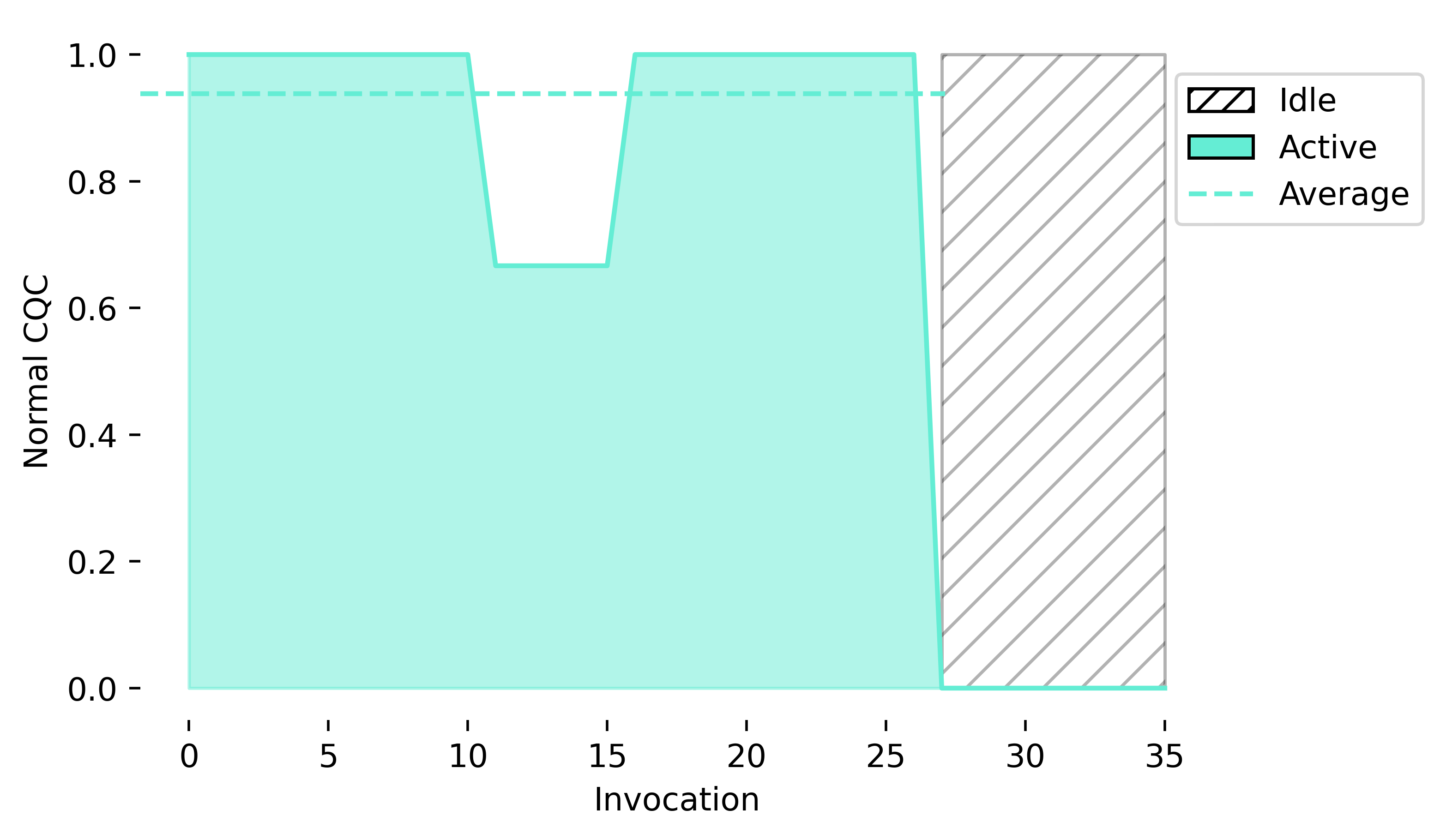}
            \caption{Worker 3}
            \label{f1_w3}
            \end{subfigure} 
 \caption{GSearch: Resource Usage in a 3-Worker Cluster Mode} 
 \label{f1}
\end{figure*}

\begin{figure*}[htbp]
\centering
         \begin{subfigure}[b]{0.32\textwidth}
            \centering
                \includegraphics[width=1\linewidth]
                {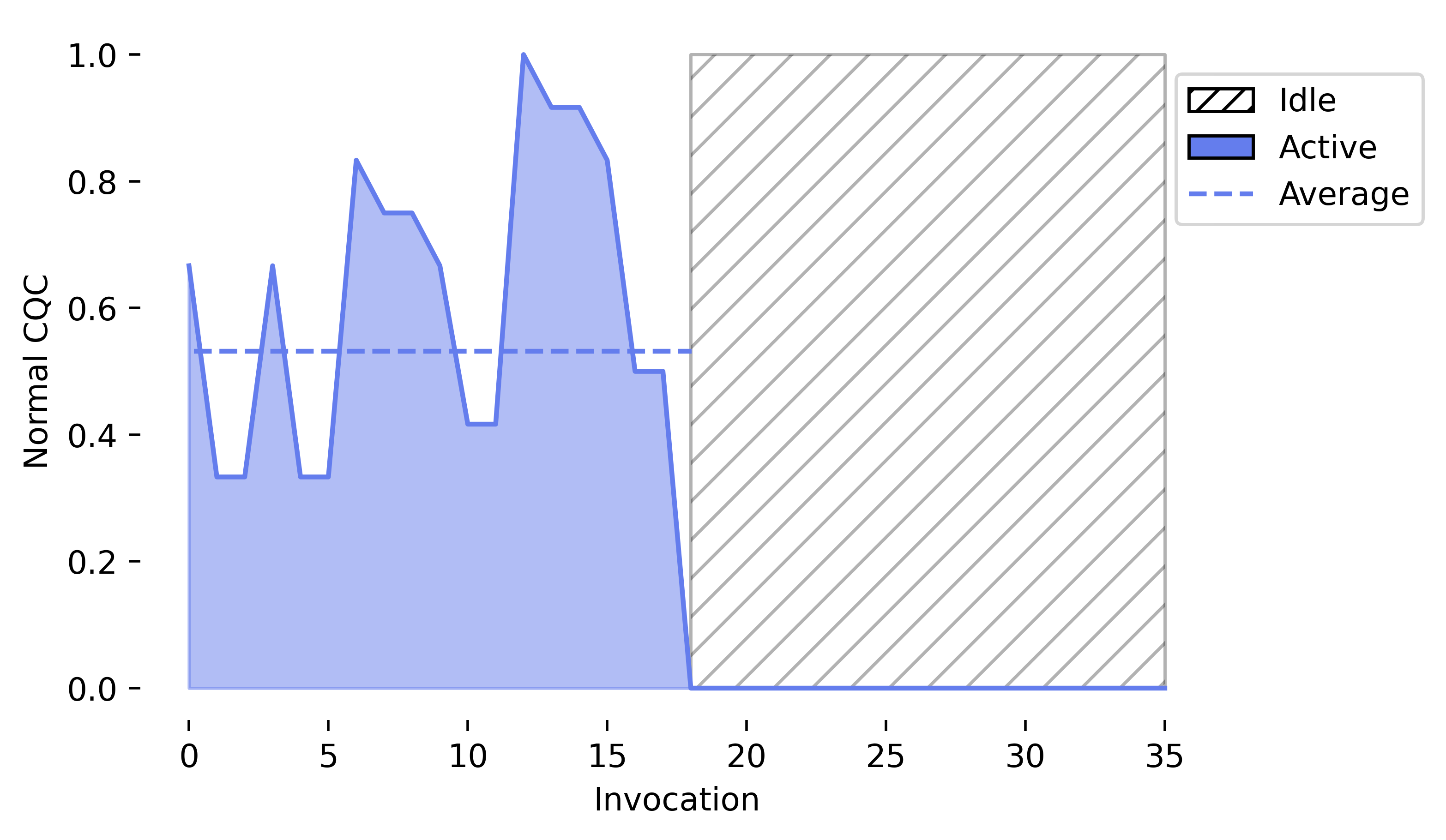}
            \caption{Worker 1}
            \label{f2_w1}
            \end{subfigure}
         \begin{subfigure}[b]{0.32\textwidth}
            \centering
                \includegraphics[width=1\linewidth]
                {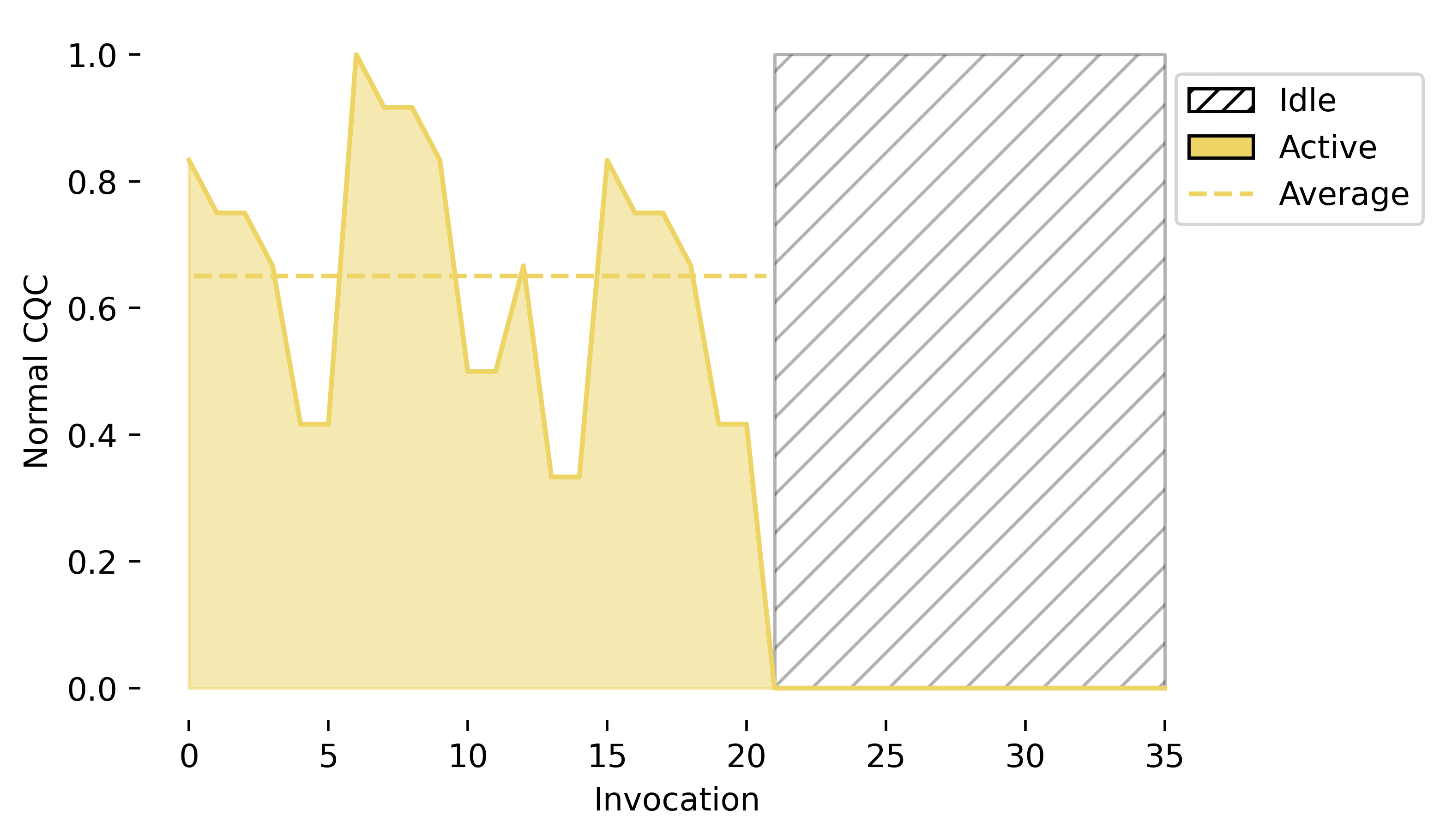}
            \caption{Worker 2}
            \label{f2_w2}
            \end{subfigure}            
         \begin{subfigure}[b]{0.32\textwidth}
            \centering
                \includegraphics[width=1\linewidth]
                {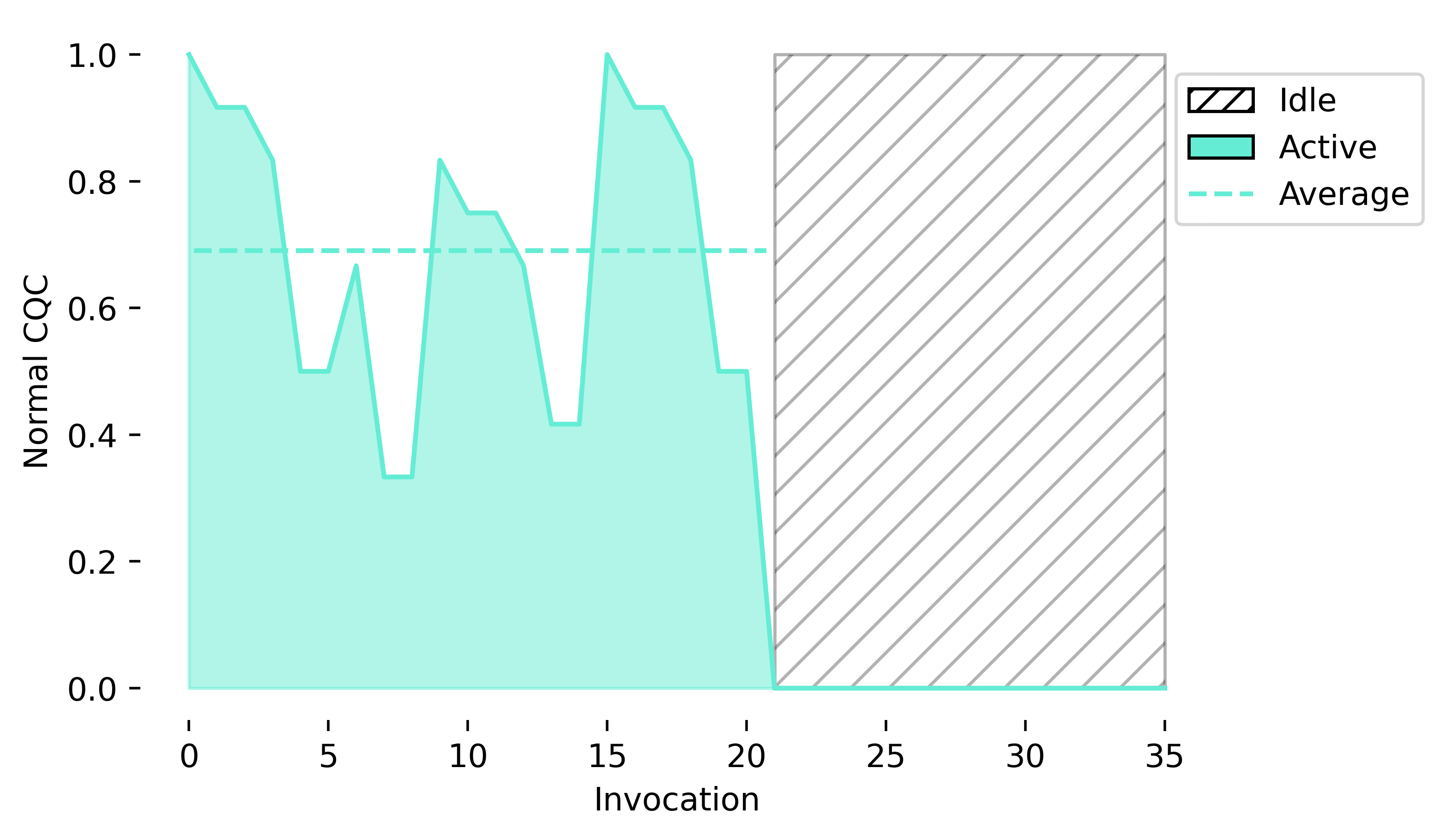}
            \caption{Worker 3}
            \label{f2_w3}
            \end{subfigure} 
 \caption{\sol: Resource Usage in a 3-Worker Cluster Mode}   
 \label{f2}
\end{figure*}

\begin{figure*}[htbp]
\centering
         \begin{subfigure}[b]{0.32\textwidth}
            \centering
                \includegraphics[width=1\linewidth]
                {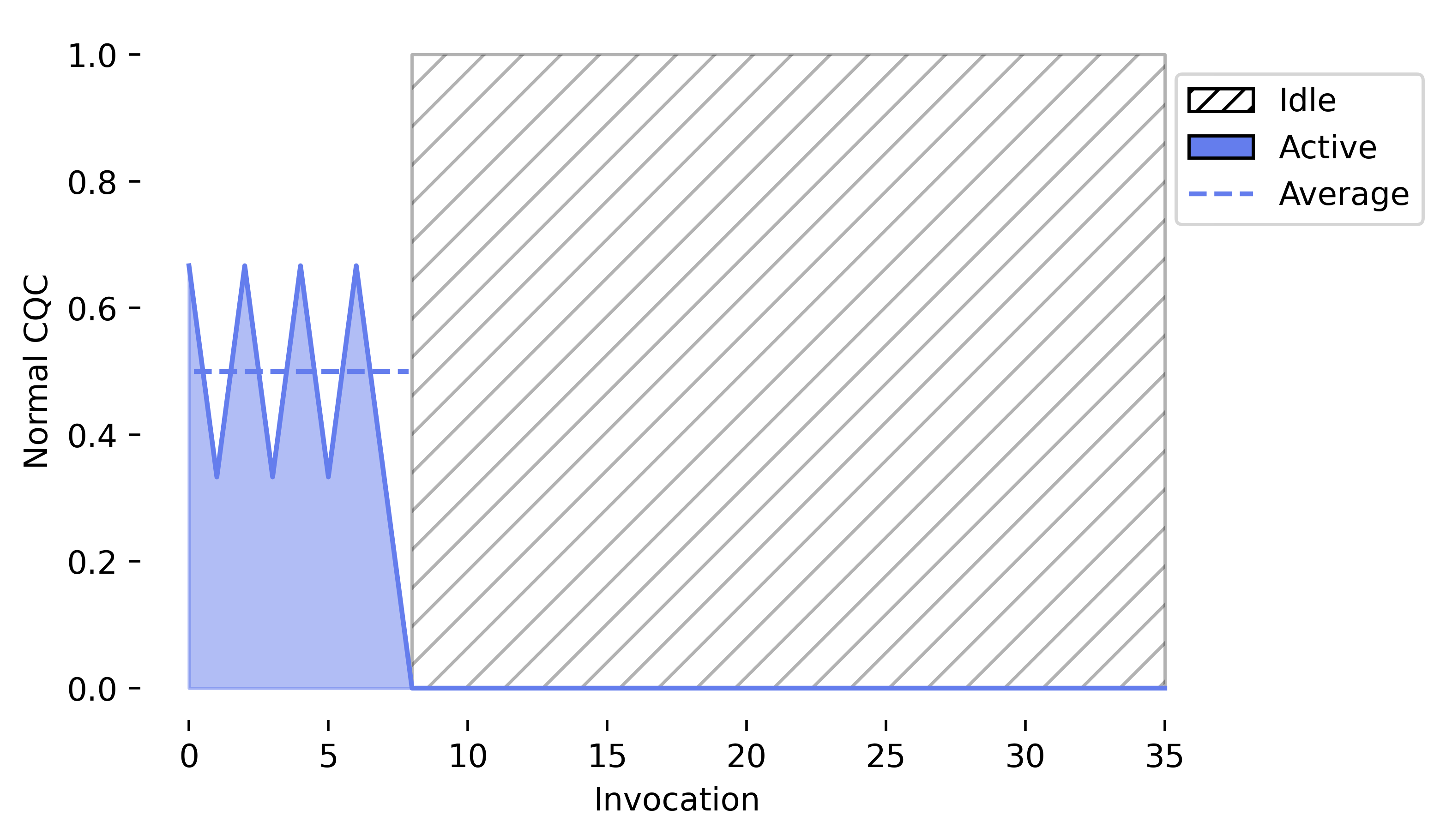}
            \caption{Worker 1}
            \label{f3_w1}
            \end{subfigure}
         \begin{subfigure}[b]{0.32\textwidth}
            \centering
                \includegraphics[width=1\linewidth]
                {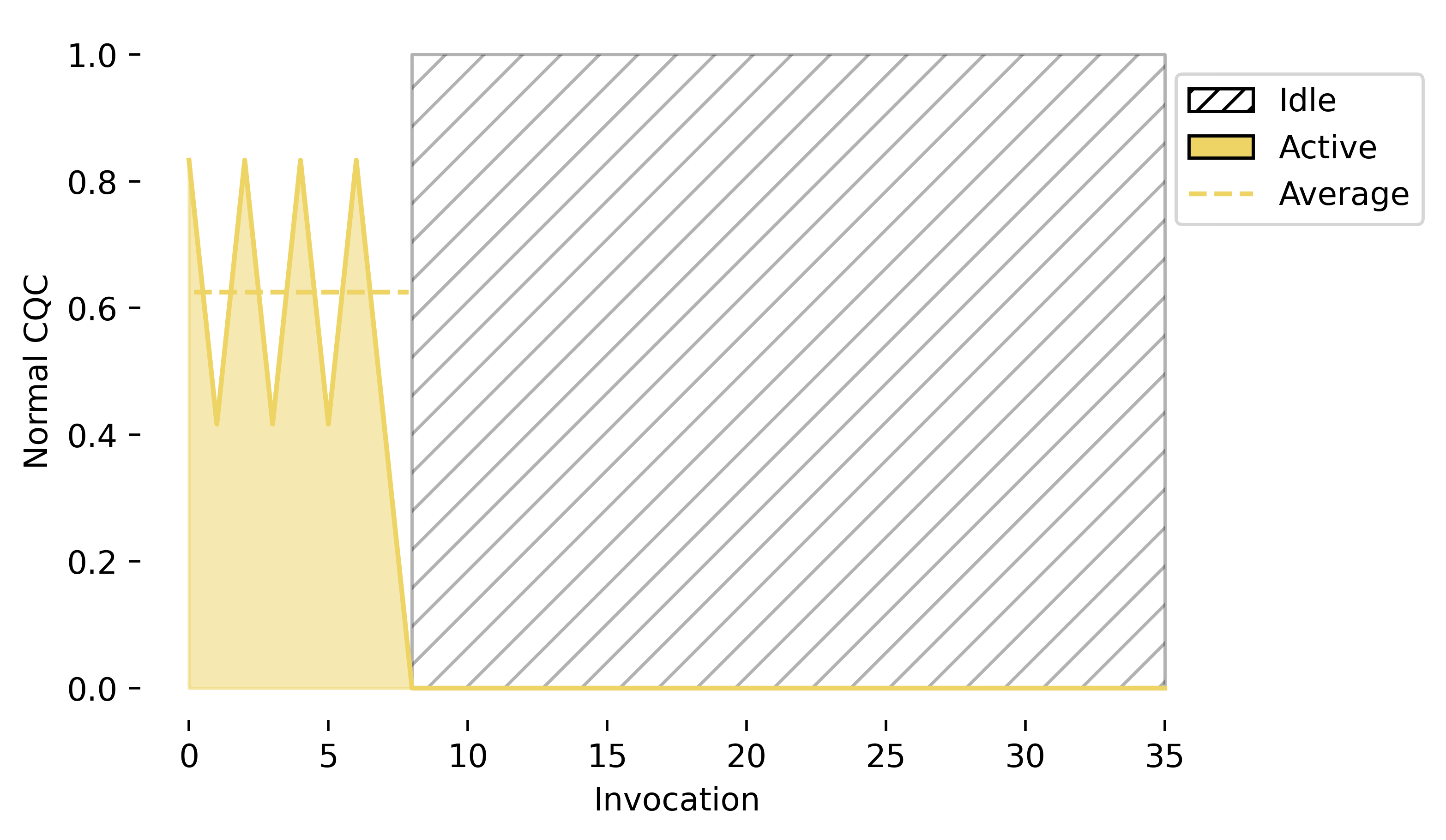}
            \caption{Worker 2}
            \label{f3_w2}
            \end{subfigure}            
         \begin{subfigure}[b]{0.32\textwidth}
            \centering
                \includegraphics[width=1\linewidth]
                {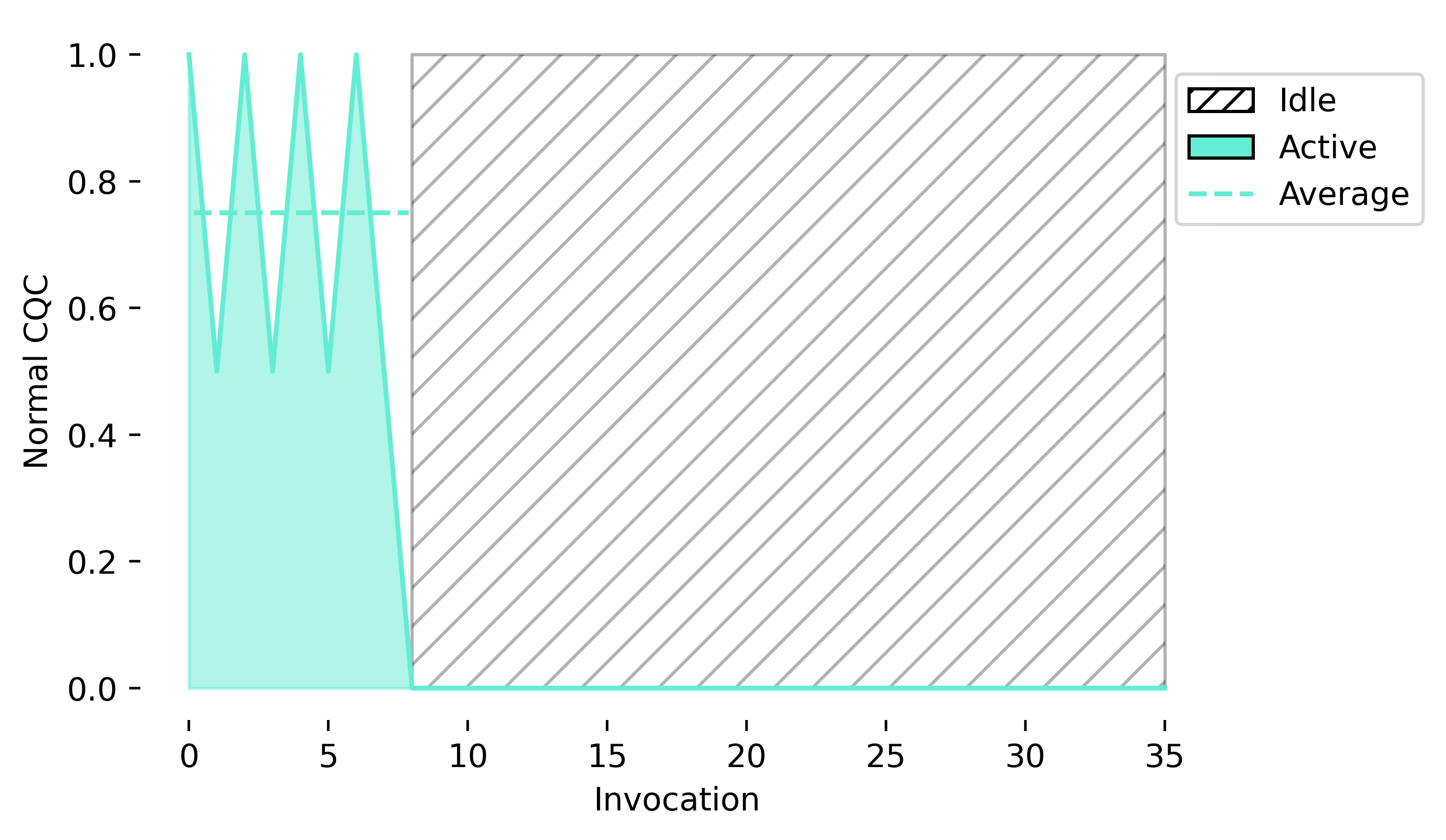}
            \caption{Worker 3}
            \label{f3_w3}
            \end{subfigure} 
 \caption{\soll: Resource Usage in a 3-Worker Cluster Mode} 
 \label{f3}
\end{figure*}

In our experiments, we submit 12 tasks to the hybrid system that contains 3 classical machines and 3 quantum workers. Specifically, we focus on resource utilization of different solutions. 
These 12 tasks are repetitions of the above-mentioned 3 different experiments, with each submitted 4 times. 
Figures~\ref{f1},\ref{f2} and \ref{f3} represent resource utilization of Gsearch, \sol, and \soll~ on 3 quantum workers, respectively, each configured with 12 qubits. The $x$-axis represents the number of invocations. 
The largest value on the $x$-axis is set to 35 because, with Gsearch, worker-2 utilizes 34 invocations which is the largest value among all experiments. The $y$-axis represents the normalized CQC values on each worker. 
The colored areas in the figures represent the active periods when qubits are occupied by searching tasks, and the shaded areas represent idle periods when resources are available. The dashed lines represent this specific worker's average normalized CQC value during active periods. 

Figure~\ref{f1} shows the results of GSearch in the 3-worker system. When processing the same task, the number of qubits remains constant with each invocation as it completes the entire search within one iteration. Therefore, the CQC value also remains constant until the task is completed. For example, in Figure~\ref{f1_w3}, during the periods of invocations 12 to 16, it represents the execution of a single task in Experiment 1. This task requires 8 qubits which is 66.67\% when normalized to the maximum value of 12 qubits. The average CQC usage across each worker is 83.95\% in worker-1, 86.27\% in worker-2, and 93.83\% in worker-3, respectively. Thus, the average CQC usage across all workers is 87.88\%. The percentage active times for the three workers are $77.14$\%, $97.14$\%, and $77.14$\%, respectively. The average percentage active time across all workers is $83.81$\%.


Figure~\ref{f2} presents the CQC using \sol. The average CQC usage across each worker is 53.17\% in worker-1, 65.08\% in worker-2, and 69.05\% in worker-3. The average CQC across all workers is 62.90\%, which is a 28.43\% reduction of the GSearch's. The percentage active times are $51.43$\%, $60.00$\%, and $60.00$\%, with an average of $57.14$\%, which is a $31.82$\% reduction of the GSearch's.

Figure~\ref{f3} presents the CQC using \soll. The average CQC usage across each worker is 50.00\%, 62.50\%, and 75.00\%, with an overall average of 62.50\%. This yields a reduction of 28.88\% compared to GSearch and a reduction of 0.64\% compared to \sol. Furthermore, the active periods for all workers amount to 22.86\%, which is a 72.72\% reduction compared to GSearch and a 60.00\% reduction compared to \sol. Therefore, using \soll effectively conserves computational resources which reduces the overall activity time. This fact is evident by comparing the shaded idle areas in Figures~\ref{f1}-\ref{f3}.

\section{Conclusion}

In this work, we present \soll, a resource-efficient and self-adaptive quantum-classical system for (index, value) unstructured search.
Based on the Grover's algorithm, it utilizes a classical postprocessing approach to analyze the resulting quantum state probabilities  iteratively. Additionally, it filters out the data points that are unlikely to be searching targets  according to the analysis. The remaining dataset is further processed into new (index, value) pairs with a mapping to the original ones. With this design, \soll~ is able to reduce the dataset iteratively.  
The reduced dataset is fed to \soll~ in the next iteration for further processing until it becomes stable. 

With classical postprocessing, \soll~ can signifcantly reduced the qubit requirement and improve the resource efficiency. 
We implement \soll~ with Qiskit and evaluate it with extensive experiments under both standalone and cluster modes. 
The results demonstrate that \soll~ reduced cumulative-qubit consumption by up to 86.36\% comparing to the state-of-the-arts. Moreover, it reduces active periods by up to 72.72\% on quanutm workers in the cluster mode.

\section{Acknowledgements}
This research was supported in part by the National Science Foundation (NSF) under grant agreements 2329020, 2301884, 2335788 and 2343535.

\bibliographystyle{IEEEtran}
\bibliography{references.bib,previous-refs.bib}

\newpage

\vskip -2\baselineskip plus -1fil
\begin{IEEEbiography}
[{\includegraphics[width=1in,height=1.25in,clip,keepaspectratio]{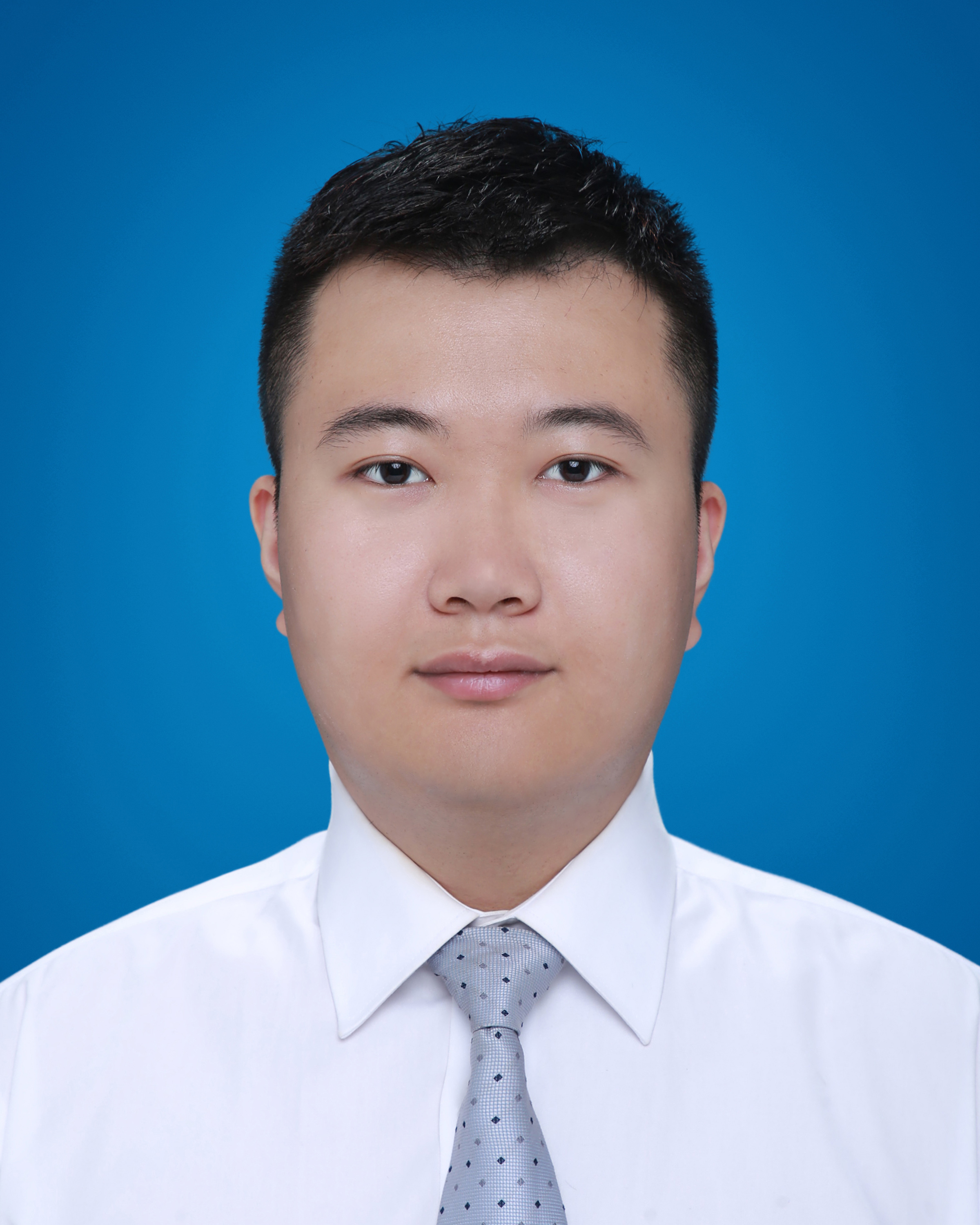}}]{Zihao Jiang}
graduated from high school in 2016 and came to the United States to continue his education. In 2021, he received the bachelor of Art in Statistics/Mathematics in Rutgers University in New Jersey. In 2022, he joined Fordham University for a master of science degree in Data Science. During his time at Fordham, he earned graduate assistantship and worked for Computer and Information Science Department. He received his Master degree and Advanced Certificate in Financial Econometrics and Data Analysis from Fordham University in 2023.
\end{IEEEbiography}
\vskip -2\baselineskip plus -1fil
\begin{IEEEbiography}
[{\includegraphics[width=1in,height=1.25in,clip,keepaspectratio]{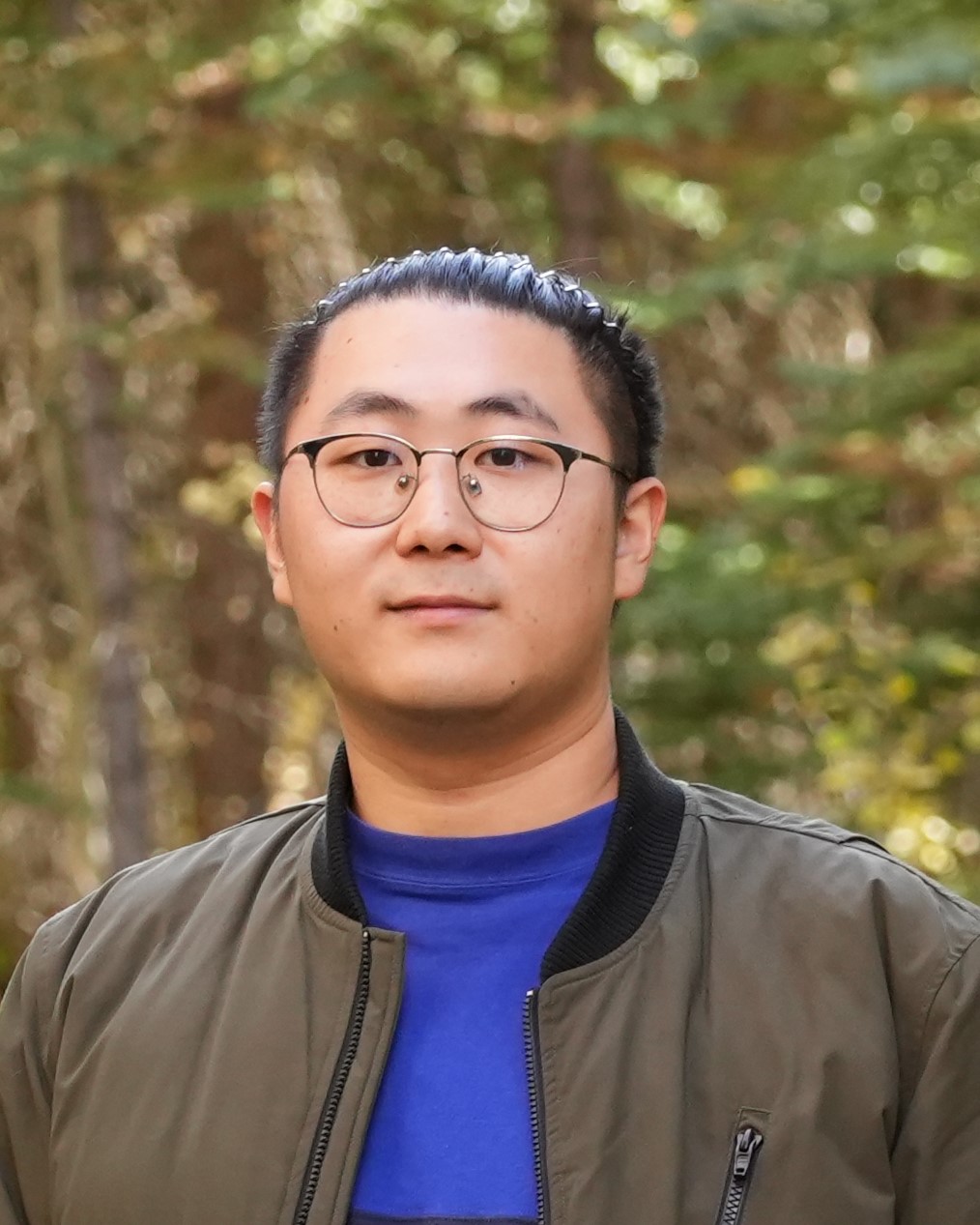}}]{Zefan Du} received the bachelor of science in Mathematics at Purdue University, West Lafayette, in 2021. He is currently pursuing his master of science degree in Data Science at Fordham University, NY. He is a graduate assistant at Fordhaam and expected to graduate in December 2023. His research interests, which include meme-stock analysis and quantum computing, reflect a diverse range of topics within the field of data science.
\end{IEEEbiography}
\vskip -2\baselineskip plus -1fil
\begin{IEEEbiography}[{\includegraphics[width=1in,height=1.25in,clip,keepaspectratio]{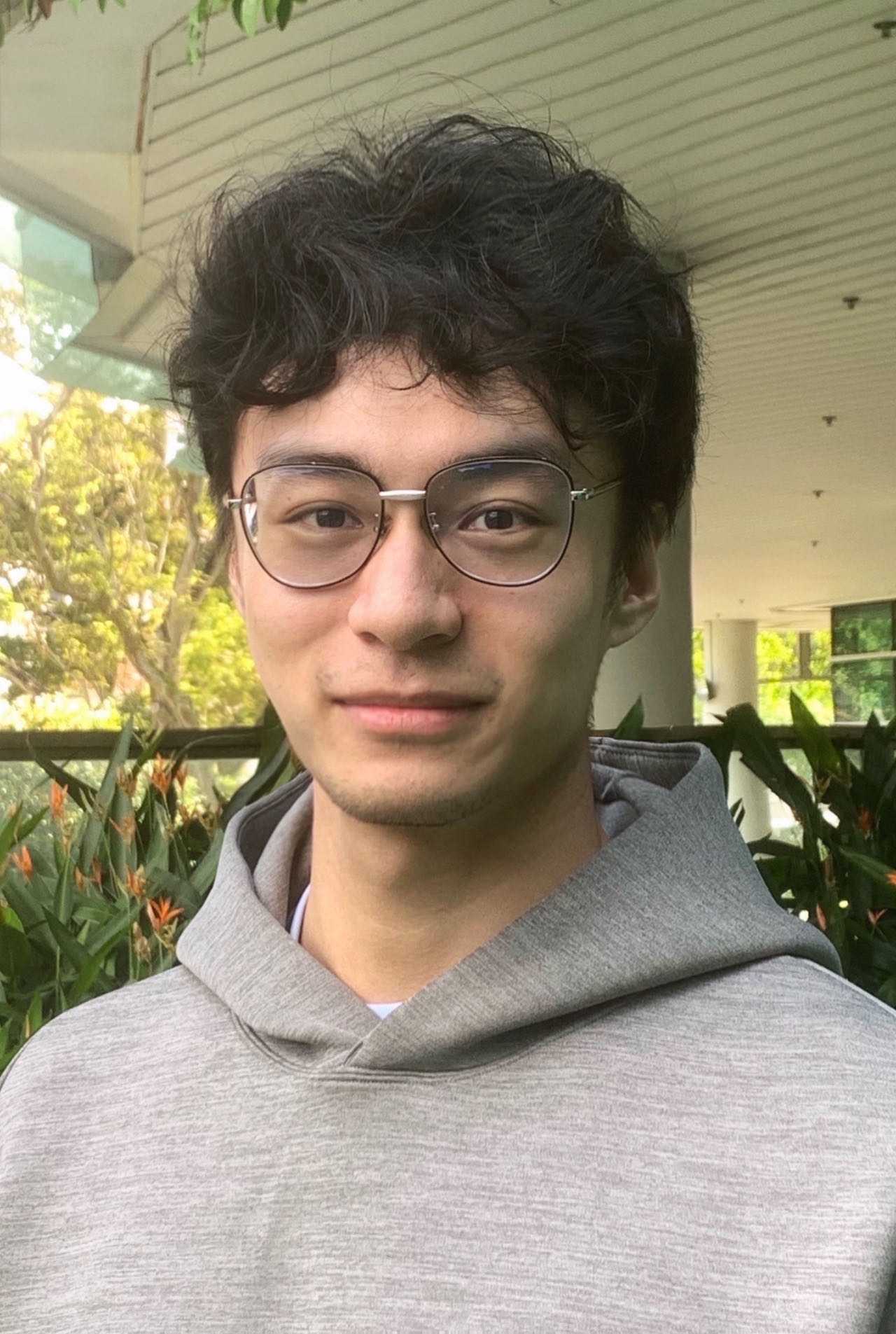}}]{Shaolun Ruan} is currently a Ph.D. candidate in School of Computing and Information Systems at Singapore Management University. He received his bachelor's degree from the University of Electronic Science and Technology of China majoring in Information Security at the School of Computer Science and Engineering in 2019. His work focuses on developing novel graphical representations that enable a more effective and smoother analysis using machines, leveraging the methods from Data Visualization and Human-computer Interaction.
\end{IEEEbiography}
\vskip -2\baselineskip plus -1fil
\begin{IEEEbiography}[{\includegraphics[width=1in,height=1.25in,clip,keepaspectratio]{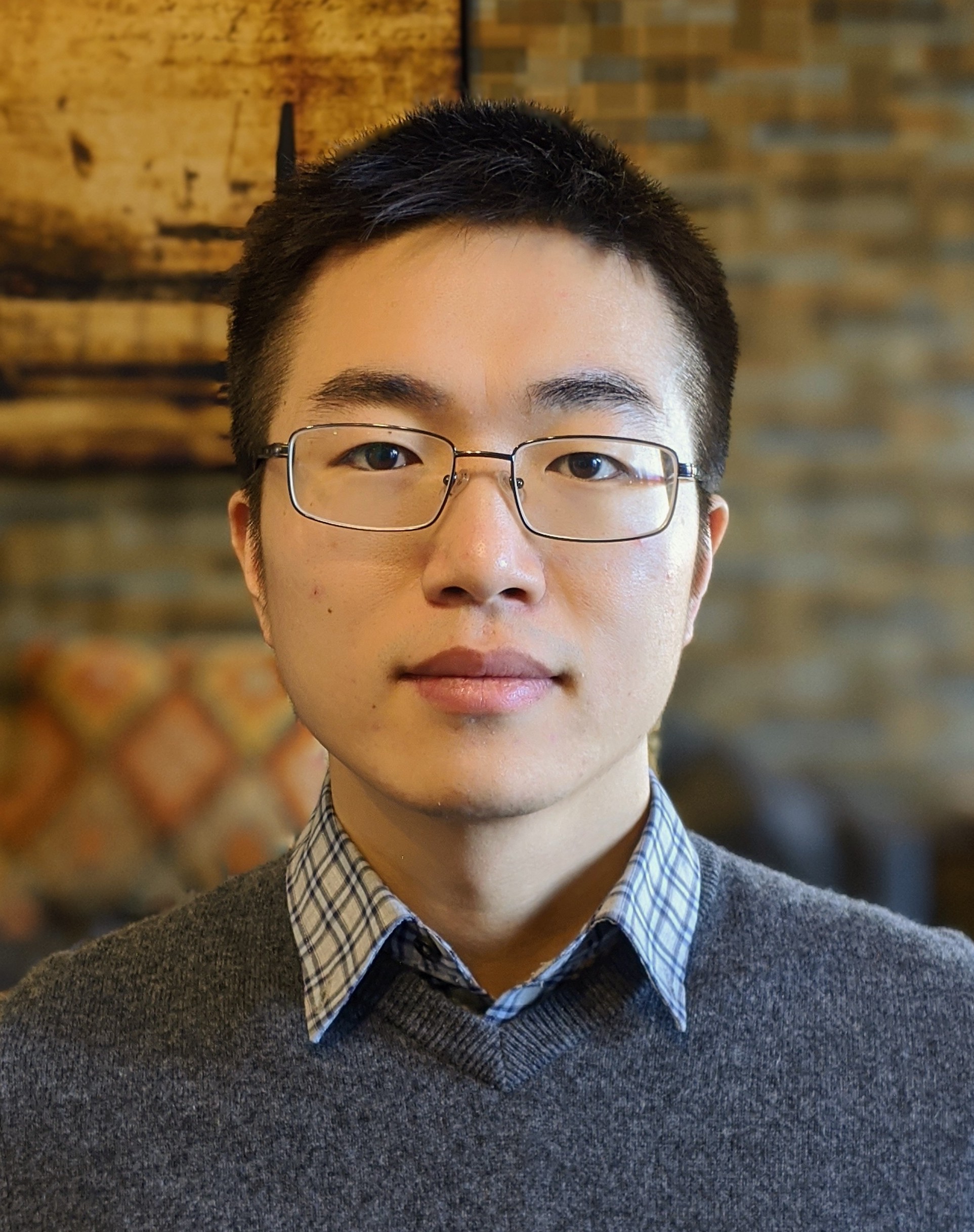}}]{Juntao Chen}
(S'15-M'21) received the Ph.D. degree in Electrical Engineering from New York University (NYU), Brooklyn, NY, in 2020, and the B.Eng. degree in Electrical Engineering and Automation with honor from Central South University, Changsha, China, in 2014. He is currently an assistant professor at the Department of Computer and Information Sciences and an affiliated faculty member with the Fordham Center of Cybersecurity, Fordham University, New York, USA. His research interests include cyber-physical security and resilience, quantum AI and its security, game and decision theory, network optimization and learning. He was a recipient of the Ernst Weber Fellowship, the Dante Youla Award, and the Alexander Hessel Award for the Best Ph.D. Dissertation in Electrical Engineering from NYU.
\end{IEEEbiography}
\vskip -2\baselineskip plus -1fil
\begin{IEEEbiography}[{\includegraphics[width=1in,height=1.25in,clip,keepaspectratio]{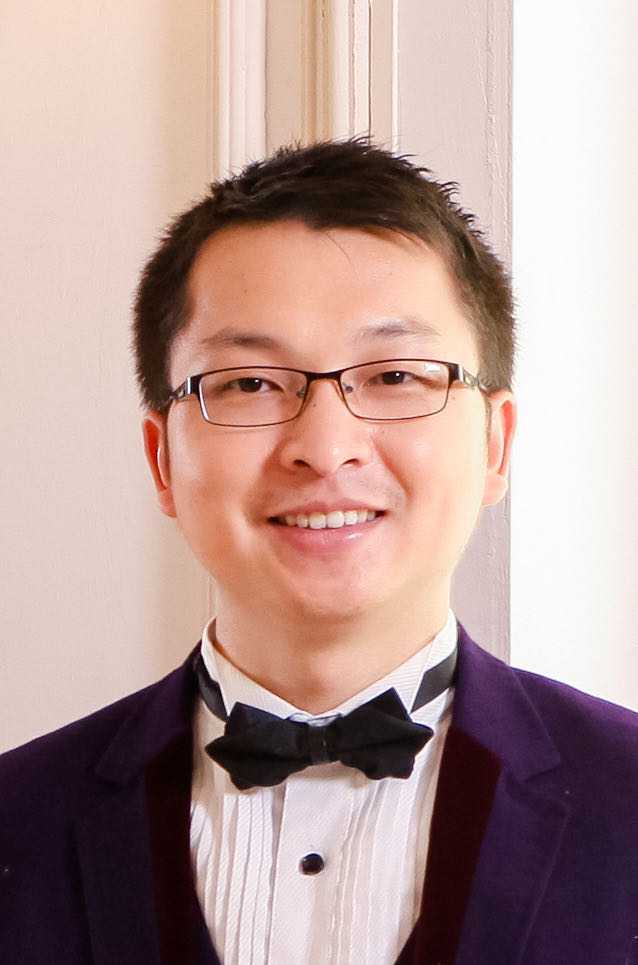}}]{Yong Wang} is currently an assistant professor in School of Computing and Information Systems at Singapore Management University. His research interests include data visualization, visual analytics and explainable machine learning.
He obtained his Ph.D. in Computer Science from Hong Kong University of Science and Technology in 2018. He received his B.E. and M.E. from Harbin Institute of Technology and Huazhong University of Science and Technology, respectively. For more details, please refer to \url{http://yong-wang.org}.
\end{IEEEbiography}
\vskip -2\baselineskip plus -1fil
\begin{IEEEbiography}[{\includegraphics[width=1in,height=1.25in,clip,keepaspectratio]{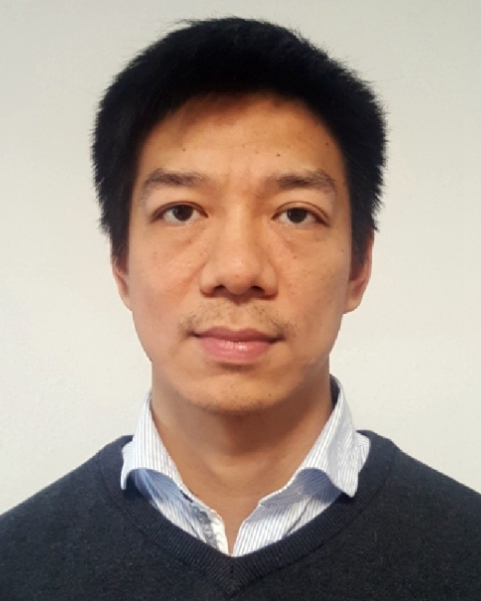}}]{Long Cheng} is a Full Professor in the School of Control and Computer Engineering at North China Electric Power University in Beijing.  He received Ph.D from National University of Ireland Maynooth in 2014. He was an Assistant Professor at Dublin City University, and a Marie Curie Fellow at University College Dublin. He also has worked at organizations such as Huawei Technologies Germany, IBM Research Dublin, TU Dresden and TU Eindhoven. He has published more than 90 papers in journals and conferences like TPDS, TON, TC, TSC,  TASE, TCAD, TVT, TCC, TBD, TITS, TSMC, TVLSI, JPDC, HPCA, CIKM, ICPP, CCGrid and Euro-Par etc. His research focuses on distributed systems,  deep learning, cloud computing and process mining. Prof Cheng is a Senior Member of the IEEE, an associate editor of IEEE Transactions on Consumer Electronics, and Co-Chair of Journal of Cloud Computing.
\end{IEEEbiography}
\vskip -2\baselineskip plus -1fil
\begin{IEEEbiography}[{\includegraphics[width=1in,height=1.25in,clip,keepaspectratio]{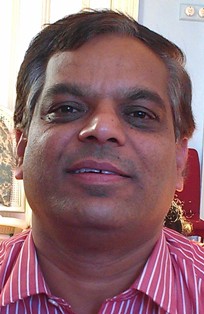}}]{Rajkumar Buyya} is a Fellow of IEEE, Academia Europea, and Redmond Barry Distinguished Professor and Director of the Cloud Computing and Distributed Systems (CLOUDS) Laboratory at the University of Melbourne, Australia. He is also serving as the founding CEO of Manjrasoft, a spin-off company of the University, commercializing its innovations in Cloud Computing. He has authored over 850 publications and seven textbooks including "Mastering Cloud Computing" published by McGraw Hill, China Machine Press, and Morgan Kaufmann for Indian, Chinese and international markets respectively. Dr. Buyya is one of the highly cited authors in computer science and software engineering worldwide (h-index=167 g-index=365, and 147,500+ citations). He has been recognised as a "Web of Science Highly Cited Researcher" for seven times since 2016, "Best of the World" twice for research fields (in Computing Systems in 2019 and Software Systems in 2021/2022/2023) as well as "Lifetime Achiever" and "Superstar of Research" in "Engineering and Computer Science" discipline twice (2019 and 2021) by the Australian Research Review.
\end{IEEEbiography}
\vskip -2\baselineskip plus -1fil
\begin{IEEEbiography}[{\includegraphics[width=1in,height=1.25in,clip,keepaspectratio]{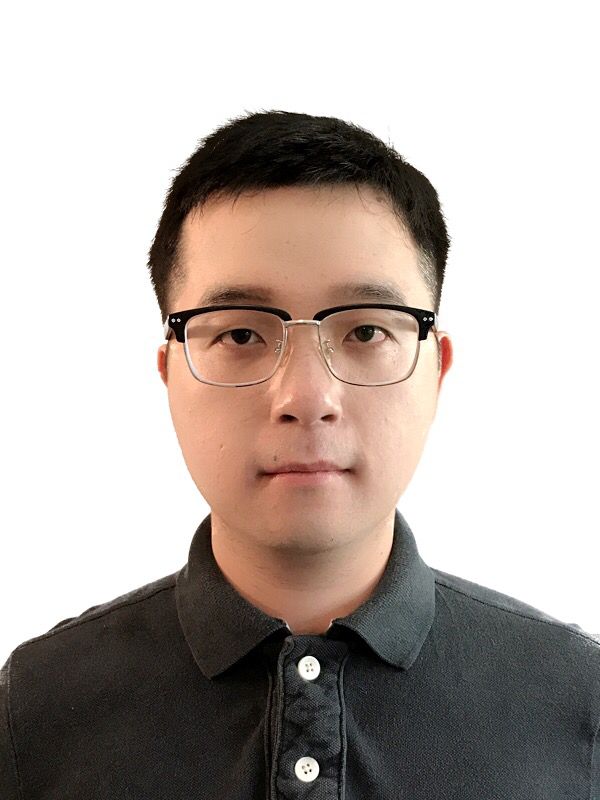}}]{Ying Mao} is an Associate Professor in the Department of Computer and Information Science at Fordham University in the New York City. He received his Ph.D. in Computer Science from the University of Massachusetts Boston in 2016. He was a Fordham-IBM research fellow. His research interests mainly focus on the fields of quantum systems, quantum deep learning, quantum-classical optimizations, system virtualization, cloud resource management, data-intensive platforms and containerized applications.
\end{IEEEbiography}


\end{document}